**Sectorial Exclusion Criteria in the Marxist Analysis of the Average Rate of Profit: The United States Case (1960-2020)**

**José Mauricio Gómez Julián**

mauriciogomezjulian@protonmail.ch

**ORCID:** 0009-0000-2412-3150

**Abstract**

The long-term estimation of the Marxist average rate of profit does not adhere to a theoretically grounded standard regarding which economic activities should or should not be included for such purposes, which is relevant because methodological non-uniformity can be a significant source of overestimation or underestimation, generating a less accurate reflection of the capital accumulation dynamics. This research aims to provide a standard Marxist decision criterion regarding the inclusion and exclusion of economic activities for the calculation of the Marxist average profit rate for the case of United States economic sectors from 1960 to 2020, based on the Marxist definition of productive labor, its location in the circuit of capital, and its relationship with the production of surplus value. Using wavelet-transformed Daubechies filters with increased symmetry, empirical mode decomposition, Hodrick-Prescott filter embedded in unobserved components model, and a wide variety of unit root tests the internal theoretical consistency of the presented criteria is evaluated. Also, the objective consistency of the theory is evaluated by a dynamic factor auto-regressive model, Principal Component Analysis via Singular Value Decomposition, and regularized Horseshoe regression. The results are consistent both theoretically and econometrically with the logic of Marx's political economy.

## I. INTRODUCTION

From the pioneering works of Shaikh and Ochoa in 1984, various research studies emerged that empirically examine the average rate of profit (ARoP) and the prices of production. The latter is a Marxist category which belongs to the microeconomic differences between prices proportional to the labor values objectified in commodities and market prices. More specifically, prices of production are the theoretical averages towards which market prices tend due to the existence of an ARoP formed through capitalist technological competition and intersectorial capital exodus.

These investigations commonly share two characteristics. First, almost all of them calculate the ARoP as the gross/net operating surplus (surplus value) over, on one hand, fixed assets and intermediate consumption (constant capital), and on the other hand, labor compensation (variable capital); this calculation methodology, which does not make a theoretical differentiation of productive and unproductive sectors to be included, and therefore leads to include all sectors, can be referred to as *naive Marxist Calculation* (nMc).

Second, share the characteristic of not theoretically defining a standard for determining which sectors should be considered in such an analysis. Therefore, the sectors included in these studies are not necessarily the same in qualitative terms. This can lead to overestimation or underestimation of the ARoP depending on which sectors are included or not, which in turn can lead to qualitatively different conclusions and have relevant theoretical consequences, such as those related to the long-term sustainability of capitalist profitability given class antagonisms. This is shown in Figure 1, in which for different calculation methodologies, a long-term decline in the ARoP is not observed.

**Figure 1**

*Gross/Net Weighted/Unweighted Average Rate of Profit for Total Economy and Corporate Non-Financial Economy*

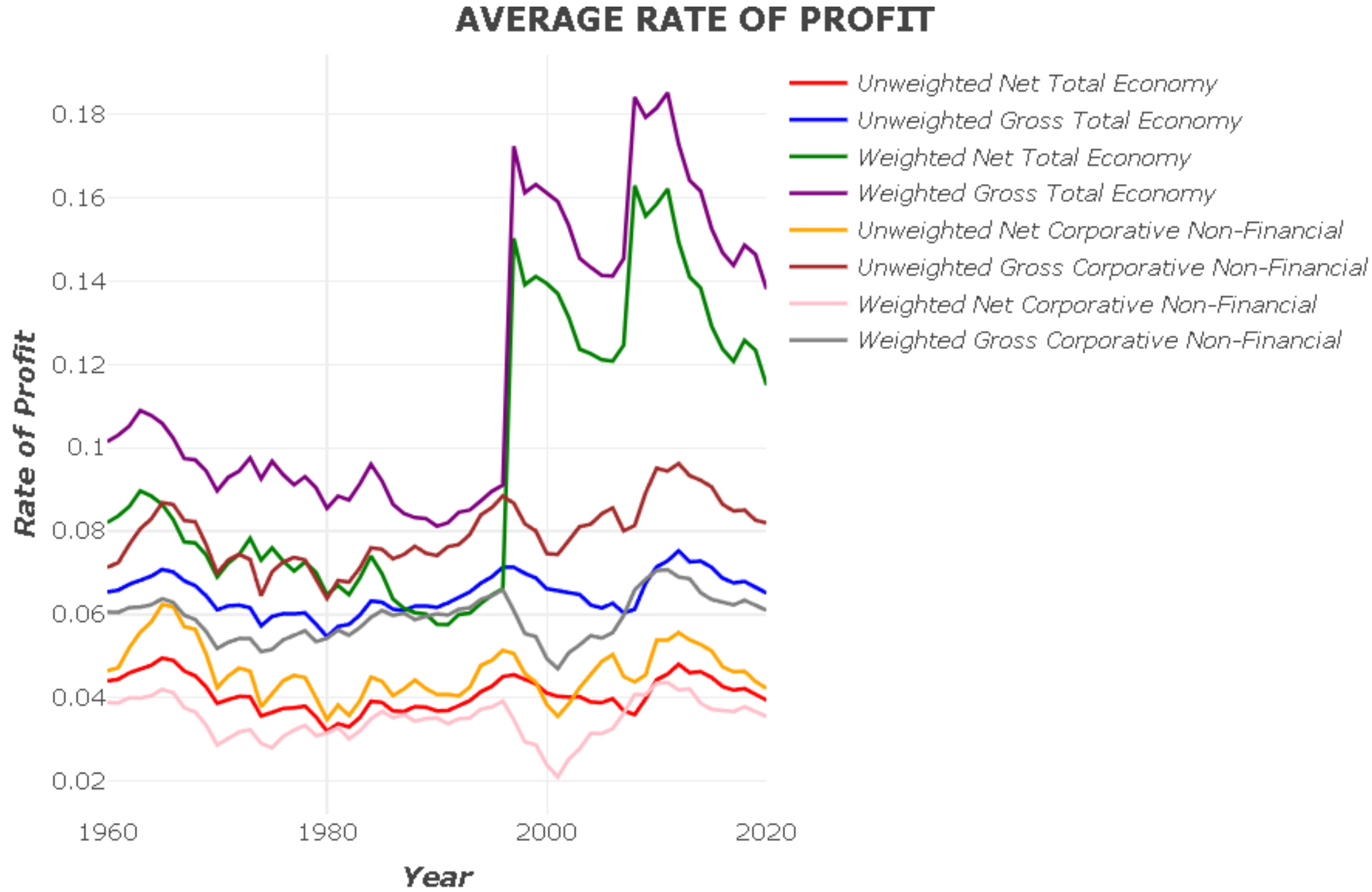

The main purpose of this research is to establish, in the light of Marx's theory, which sectors should be taken into consideration when studying the long-term ARoP and why this should be done in such a manner. This will be done by considering the theoretical differences between productive and unproductive labor[1], the sector location in the complete circuit of capital as a whole and the sectors link with the production of surplus value.

The relevance of the set of criteria in question lies in ensures methodological uniformity among investigations of the same kind, which is relevant in terms of the reproducibility required by the scientific method, and in allows for obtaining standardized estimations of the ARoP that truly govern the process of surplus value accumulation by capital, thereby eliminating a significant methodological source of overestimation and underestimation[2].

As stated by (Young 2018), the truth coherence theory states that a proposition is considered true if it is in harmony or agreement with other statements in a specific set. This is the internal consistency of the statement in question. So, suppose we have a group of statements, and these statements must fit well together, so that there are no contradictions or logical conflicts among them. If a new statement fits perfectly with the existing set of statements without creating contradictions, then it is considered true according to a truth coherence theory.

Consequently, since it is well known that Marx purposes that the long-term average rate of profit tends to decline, if the statistical estimation of this rate shows a downward trend, the obtained results will serve as strong evidence of the internal consistency of the proposed criteria, which would play the role of the new proposition introduced into the system of propositions. To achieve this, the analysis will utilize the three most modern filters in the analysis of economic time series and four types of unit root tests with their variations (a total of 17-unit root tests).

However, it is important to note that the internal consistency of a theory does not guarantee its overall validity of the theory. Therefore, this would be an important but not definitive step in the evaluation of Marxist theory. Therefore, an additional step will be taken, now in the evaluation of the external or objective consistency of the theory, that is, in relation to its strength as an explanation of objective reality, which it going to be done by the implementation of Principal Component Analysis (PCA), Singular Value Decomposition (SVD), regularized Horseshoe regression (RHR), and a dynamic factor auto-regressive model (DFM).

The complementary implementation of those models will allow us to verify from the perspective of various methodologies, the correspondence between the resulting production matrix based on the developed criteria (theoretically presented in Section II.II, and whose applied results to US economy for 1960-2020 will be presented in Section III.I) and the relevant production matrix obtained through the PCA, BW, and DFM methodologies.

[1] Productive and non-productive sectors are those in which the workforce performs productive and unproductive tasks, respectively. Therefore, referring to one or the other, in cases where they are not hybrid sectors (*i.e.*, with productive and unproductive labor), is equivalent. We will define *productive labor* in Section II.II.

[2] Since theoretical errors lead to methodological errors.

## II. METHODOLOGY

### II.I. GENERALITIES

This research is composed of two parts. The first part involves constructing, based on Marxist theory, the criteria for sectoral inclusion, starting from the connection between economic sectors and the process of surplus value production and the circuit of capital. The second part aims to demonstrate the internal theoretical consistency of the presented criteria. This consistency is established in terms of the Marxist theory's assertion that, in the long run, there is a decreasing tendency of the ARoP of productive capital, and consequently, the ARoP calculated with sectors classified as productive should tend to decrease. To achieve this, three types of time series decomposition are employed to extract the trend of the variable in question, which are the less asymmetric Daubechies wavelet transformation for multiresolution analysis[3], empirical mode decomposition to capture intrinsic non-linearities, and the Hodrick-Prescott filter embedded in an unobserved components model.

### II.II. BULDING EXCLUSION CRITERIA

It is advisable to study separately the productive capitalist sector from the non-capitalist sector (those in which the purpose of production is not capital accumulation, *i.e.*, they do not produce with the aim of maximizing profit) and the non-productive sector (which includes economic branches that solely redistribute surplus value and do not produce it, specifically wholesale and retail trade, finance, real estate, and government sectors) (Cheng & Li, 2020, p. 117).

On the other hand, it should be noted that Guerrero Jiménez (2000, pp. 97-100) excludes the following sectors: "General services of public administrations", "Non-commercial research and education", "Non-commercial healthcare", "Non-commercial services not elsewhere classified", "Real estate rental" (due to the inclusion of a component that he considers fictitious, which is self-renting), "Imputed production of banking services", and pure circulation work (which Guerrero does not exclude). He also points out that the Marxist tradition excludes the commercial and financial sectors (although Guerrero includes a sector called "Credits and insurance").

On the other hand, Ochoa (Guerrero Jiménez, 2000, p. 114), a pioneer in the statistical research of prices of production from the perspective of the simultaneous school, does not exclude the sectors related to pure circulation (sector 61) or the government

[3] A technique that decomposes the signal into distinct frequency bands across varying time scales, thereby enabling the simultaneous isolation of long-term trends and transient shocks without sacrificing temporal localization.

sector (sector 64), as well as the trade sector (sector 65) and the financial sector (sector 66).

Why should pure circulation sectors (commercial, financial and some services sectors), and government be excluded when analyzing the ARoP? The reasons are as follows:

1. Both the commercial and financial sectors are not value-creating sectors, but rather sectors that redistribute the value created by productive sectors. Regarding pure circulation, Marx states (Marx, El Capital 2010c, p. 276) that "We have already seen in Book II [pp. 108-111] [4] that the mere functions performed by capital in the sphere of circulation - the operations that the industrial capitalist must carry out, firstly to realize the value of commodities and secondly to convert this value back into the elements of production of the commodity, the operations necessary to serve as the vehicle for the metamorphoses of commodity-capital M'-D-M, that is, the acts of sale and purchase - do not create value or surplus value."
2. The government sector is not oriented towards profit maximization in its production and therefore does not correspond to the real dynamics of capitalist competition, which is driven by the pursuit of profit maximization.
3. Regarding services, not all services are value-creating. Which services create value? This will be addressed next, and the resulting classification of services based on whether they produce surplus value or not will be justified precisely by the sectoral inclusion criteria presented in this research.

The question regarding economic activities related to services does not seem to have been sufficiently explored in Marxist theory. When Marx wrote his work, the concept of the "service sector" did not exist in macroeconomic statistics; in fact, the concept of "macroeconomic statistics" itself did not exist as such. Nor did the concept of "sectors" exist as it is currently understood, that is, as a theoretical-statistical framework for classifying agricultural, industrial, or service activities. The terms Marx used were "productive branches" (which would correspond to modern-day input-output industries) and the category of "sectors" in his thinking refers to the "sector producing means of production" and the "sector producing means of consumption".

---

[4] In the passage pointed out by Marx, he writes, "Within the sphere of circulation, capital assumes the forms of commodity-capital and money-capital. Its two processes of circulation consist of transforming from the commodity form into the money form and from the money form into the commodity form. The fact that the transformation of the commodity into money represents, at the same time, the realization of the surplus value embodied in the commodity, and that the transformation of money into commodities implies, in turn, the transformation or reversal of capital to regain the form of its elements of production, does not in the least alter the reality that these processes, as circulatory processes, are processes of mere metamorphosis of commodities."(Marx, 2010b, pp. 110-111).

The theoretical difficulties associated with the decision to include or exclude the service sector become evident upon detailed examination. Consider, for instance, the transportation of an already-produced commodity to its final consumer or to those who will use it as an intermediate input, as well as the costs of circulation arising solely from changes in the forms of commodities.

As Marx (2010b, pp. 132-133) stated:

> The general law is that *all costs of circulation, which arise only from changes in the forms of commodities do not add to their value* (...) But the use-value of things is materialised only in their consumption, and their consumption may necessitate a change of location of these things, hence may require an additional process of production, in the transport industry. The productive capital invested in this industry imparts value to the transported products, partly by transferring value from the means of transportation, partly by adding value through the labour performed in transport. This last-named increment of value consists, as it does in all capitalist production, of a replacement of wages and of surplus-value (...) The transport of products from one productive establishment to another is furthermore followed by the passage of the finished products from the sphere of production to that of consumption. The product is not ready for consumption until it has completed these movements (…) the general law of commodity production holds: The productivity of labour is inversely proportional to the value created by it. This is true of the transport industry as well as of any other. The smaller the amount of dead and living labour required for the transportation of commodities over a certain distance, the greater the productive power of labour, and vice versa.

In short, what Marx is pointing out is that it is essential to distinguish between two fundamentally different types of circulation costs: on the one hand, pure circulation—comprising mere changes in form without material transformation, such as buying and selling operations, accounting procedures, and the handling of legal titles—remains unproductive, as it adds no value to commodities and must be covered as *faux frais* deducted from surplus-value; on the other hand, transportation—which involves a material change of location—constitutes productive activity that genuinely adds value to commodities through the depreciation and consumption of means of transport, the wages paid to transport workers, and the surplus-value generated in the process, even though it may outwardly appear as a mere cost of circulation[5].

Also, does a domestic worker create value? And what can be said about women who perform domestic labor without receiving a salary (but receive support from their husbands)? And what about professional services? Should they all be considered in the same way? What is clear is that the Marxist analysis of services (and any productive sector or economic activity) must be approached in terms of (Tregenna, 2009, p. 1):

1. Their location is within the circuit of capital[6].
2. Their relationship to the production of surplus value.

It is also clear that the "service sector" includes activities that are very heterogeneous in terms of the two previous points, because (Tregenna, 2009, p. 1):

1. There are activities in which surplus value is directly produced.
2. Activities that facilitate the production of surplus value elsewhere (or increase the rate at which it is produced).
3. Activities that remain outside the circuit of capital.

The above description is an important step as it considers not only economic activities that directly produce surplus value but also those activities that, although not directly extracting it, are a necessary condition for such extraction. To properly define which complementary economic activities are indispensable for these purposes, it is necessary to adequately define two things: surplus value and the difference between productive and unproductive consumption. Surplus value is the value that unpaid labor of the wage worker (which involves a relationship of subordination of labor to capital, whether formal or real[7]) creates beyond the value of their labor

---

[5] From the standpoint of the individual industrial capitalist, outlays for transporting the finished commodity—whether paid directly to a carrier or indirectly via a merchant's margin—appear as circulation costs that reduce the seller's realized profit. At the level of social capital, however, these outlays remunerate a distinct productive branch—the transport industry—which transfers value from means of transport and adds new value through living labour (wages plus surplus-value). By contrast, the merchant's own buying-and-selling functions (invoicing, accounting, etc.) are pure faux frais and do not create value.

[6] By "circuit" we refer to the cycle that capital goes through, starting with commodities filled with surplus value (*i.e.*, capital-commodities, the commodities that have just resulted from the previous production period), transforming into money-capital in the market (it is money-capital and not just money because it is money intended for the renewal and expansion of capital in order to continue extracting surplus value), and then transforming back into capital-commodities to renew the production cycle.

[7] "The production of absolute surplus value is the general foundation on which the capitalist system rests and the starting point to produce relative surplus value (...) The production of relative surplus value presupposes a *specifically capitalist mode of production*, which can only arise and develop with its own methods, means, and conditions, through a natural and spontaneous process

power[8] and which is appropriated freely by the capitalist, *i.e.*, the owner of the means of production that take on the concrete-historical form of capital.

The above is consistent with what was discussed by (Watanabe, 1991, p. 35), who points out that Marx's concept of service fundamentally differs from the conventional (neoclassical) approach, which some Marxist economists also adopt; this is also noted by (Gao & Watanabe, 2023, p. 59). The conventional argument is based on the idea that a service society is simply characterized by having most of its jobs in service sectors, but these sectors are diverse and heterogeneous in terms of their socio-economic functions (Watanabe, 1991, p. 35). Marx advocates for a radical critique of this conventional conception and argues that we must disaggregate and functionally define activities within these sectors to understand their true nature (Watanabe, 1991, p. 35).

Therefore, even though theoretical evidence implies that services, understood as tertiary activities (those not immediately exchanged for capital but for income, with the expectation of producing surplus value rather than use value, *i.e.*, labor in the immediate link before final consumption or that provides the suitable means for final consumption), are not productive labor activities (Marx, 1969, pp. 403-404), (Watanabe 1991, pp. 26-28), (Gao & Watanabe 2023, p. 34), it is necessary, as will be shown in Section III.I, not to a priori exclude services. Instead, we should consider their internal composition based on the criteria outlined above. Based on this composition, we can determine, considering the criteria presented in this section, whether they consist of productive labor, unproductive labor, or are mixed. In the latter case, we should determine (if possible) the proportions of their mixed composition and based on this decide whether they should ultimately be considered productive labor or not.

On the other hand, the difference between productive and unproductive consumption should be understood as follows. Productive consumption directly integrates into the process of production and implies the utilization of various means of production (machines, tools, fuel, raw materials, materials, etc.) in that process. In contrast, unproductive or personal consumption refers to the consumption performed by individuals to satisfy their needs using various products (food items, clothing, footwear, consumer goods, etc.) (Borisov, Zhamin & Makarova 2009, pp. 46-47). (Marx, 1975, p. 223) states that unproductive consumption refers to any consumption that does not aim to produce something, which could serve as an equivalent to

---

based on the formal subordination of labor to capital. This formal subordination is replaced by the real subordination of the worker to the capitalist. It is enough to mention the intermediate forms in which surplus value is not extracted from the producer by direct coercion, nor does it arise from the formal subordination of the worker to capital." (Marx, 2010a, pp. 426-427), as is the case with independent producers, informal trade, etc.

[8] The value of labor power is determined by the value of the means of subsistence necessary for the survival of the worker and their family members, for the reproduction of labor power.

it. In other words, productive consumption is itself a mean, a mean for production, whereas unproductive consumption is not a mean but an end. The enjoyment involved in unproductive consumption is the good that constituted the motive for all the operations preceding it. In productive consumption, nothing is lost, whereas everything consumed unproductively is lost. Marx highlights that what is consumed productively is always capital, which is a characteristic of productive consumption that deserves particular emphasis. In the mentioned passage, Marx points out that everything consumed productively becomes capital and is transformed into capital. Furthermore, he states that the two types of consumption correspond to the two types of labor, productive and unproductive.

Regarding the consideration of capital turnover periods (Marx, 2010c, p. 152) states that the difference between profit rates (assuming equal rates of surplus value and keeping all other factors constant) results from differences in the turnover periods of employed capitals and the value proportion between the organic components of capital in different branches of production. (Marx, 2010c, p. 150) had previously noted that Adam Smith showed how profit rates tend to equalize either due to real factors[9] or cultural factors (Gómez Julián, 2022, pp. 26-27). Therefore, it is not necessary to explicitly consider either the differences in turnover periods or the differences in the degree of labor exploitation (rate of surplus value), because the former is implicitly taken into account through the estimation of weighted average sectoral profit rates (as profit differentials exist to the extent that differences in organic composition and turnover periods exist), while the latter tends towards its average in the long run. However, (Cheng & Li, 2020, p. 116) point out that in the short term (their research considers only one period), not considering turnover periods leads to "unrealistically high" profit rates, but they do not mention anything that alters its long-term trend, or the qualitative results derived from the quantitative results[10].

Additionally, (Marx, 2010c, pp. 152) states that when referring to the organic composition or turnover period of capital in a particular branch, it always refers to the weighted average normal proportion of capital invested in this branch of production and not to the fortuitous differences[11] of the different capitals invested in that branch.

One way to statistically evaluate the consistency of inclusion and exclusion criteria in terms of the theory under which they were designed is, since it establishes the fall of the ARoP as a long-term trend, to analyze whether the trend in question for the selected sample is indeed decreasing or not. To do this, filtering methods will be used.

---

[9] Linked to the economic dynamics of capitalist competition in a strict sense.

[10] For example, which sectors appropriate more surplus value than others through rate differentials, the short and long-term trends of such rates, etc. Different quantitative results can lead to the same quality, since the overestimation of short-term profit rates caused by not considering turnover periods occurs for all economic activities, and the authors do not indicate that this phenomenon occurs more strongly in one activity or another.

[11] Here, it refers to casual differences, that is, non-essential ones. The definition of the casual and its relationship with necessity (what has the force of law) can be found in (Nabi 2022, 144-147).

The criteria outlined above, which are evidently much more complex than a cooking recipe and therefore are not presented as one would with a list of ingredients, serve as the general guidelines leading to the results which are going to be presented in Section III. These results consist of analyzing the qualitative breakdown of U.S. economic activities (sectors) based on input-output information provided by the Bureau of Economic Analysis (BEA) and thus determining which activities should be included in the calculation of the Marxist ARoP and which should not. In cases where there is no possibility of ambiguity, the results will be presented directly, as would be the case for including the wood production sector or excluding the finance and insurance sector. However, for those cases in which economic activities are hybrid, meaning that some of them correspond to productive labor and others to unproductive labor, a specific analysis will be provided.

### II.III. ON THE NECESSITY OF STATISTICAL FILTERS IN THEORETICAL AND APPLIED RESEARCH

As mentioned by Bisht & Ram (2022, p.83), most of the signals we work with include data that is slightly erroneous or contains some unwanted signals or "noise". To carry out signal processing and analysis, it is imperative to remove such interference or at least reduce their effects. To extract information from noisy data, we apply filtering techniques. The goal of filtering is to estimate the real state and/or predict the future value. A filter is a system that allows certain frequencies to pass (or amplify), while reducing (attenuating) other frequencies. It discards (ignores) the undesired frequencies of the signals and thus helps extract the desired frequencies. This allows separating time series into specific components of interest, which for this research is the trend component.

This research will apply three filtering methods: the Daubechies wavelet transformation with increased symmetry (DW) (Daubechies, 1999, pp. 6, 168, 194-198), (Daubechies, 1988, p. 920), the empirical mode decomposition (EMD) (Zeiler et al., 2010, p. 1), (Huang et al., 1998, pp. 933-948), and the Hodrick-Prescott filter embedded in unobserved variables/components model or Embedded Hodrick-Prescott (EHP) (Grant & Chan, 2016, pp. 114-115). All implemented filters perform an additive decomposition. An additive decomposition of the time series is suitable for series with relatively constant seasonal variation over time, while a multiplicative decomposition is more suitable for series with increasing seasonal variation over time (Penn State Eberly College of Science, 2022).

To statistically determine whether the seasonal variation of a specific time series is relatively constant or not, a length of consecutive sub-periods must first be chosen for conducting a seasonality analysis. According to (Burns & Mitchell, 1964, pp. 440-442), a Kondratieff wave corresponds to 54-60 years (and to six Juglar cycles), a Juglar wave corresponds to 9-10 years, and a Kitchin wave corresponds to less than 40 months. Since the analyzed sample consists of 61 observations, it is not possible to study the seasonality of sub-periods that align with Kondratieff waves, as the database is based on years and does not allow for fractions of a year. Additionally, it is highly arbitrary to choose which of the possible whole annual periods, smaller than 40 months, should be selected (one, two, or three years). Therefore, the seasonality analysis will be conducted with the maximum number of Juglar waves into which the period 1960-2020 can be divided.

Finally, I consider that there are epistemological reasons related to the need to use filters in the investigation of long-term trends in dynamic systems like the capitalist mode of production. These reasons are based on the philosophical definitions that Marxist theory makes of essence and phenomenon (Rosental & Iudin, 1971, pp. 147-148), which imply that filters are necessary to reveal trends that are hidden when mixed with seasonality, cycle dynamics, and random shocks. This randomness must be understood as the fundamental property of mass random events consisting of system behaviors because of the influence of contingent variables (Rosental & Iudin, 1971, pp. 454-455).

## III. RESULTS

### III.I. EXCLUSION CRITERIA

If we analyze long-term data on US economic activity provided by the Bureau of Economic Analysis (BEA), specifically for the period between 1960 and 2020, we find the existence of three lists that enumerate such activities for the given period[12], hence it is necessary to consolidate the lists and, based on the consolidated list, define the sectors that will be excluded from the calculation of the Marxist ARoP.

As observed, for the case of the United States in the specified period (1960-2020), there are significant changes in the activities that compose its economy, because of the expansion of its material base or, in other terms, due to the development of the productive forces of labor in the nation. The first set of industries exists from 1960 to 1962 data, then from 1963 to 1996 the industries change, and again from 1997 to 2020. Thus, the first step to be taken is to express all economic activities from each period in terms of the economic activities of the first period, that is, aligning all classifications with the first available classification.

The consolidated classification table can be reformulated in such a way that on the right side of the economic activities that should be included (which directly generate surplus value and/or are an indispensable condition for its generation[13], as well as involve productive consumption), the number 1 is written, and in the opposite case, the number 0. The results are presented below.

[12] The complete dataset and its methodology documentation are in (Gómez Julián 2023).
[13] It should not be confused that its indispensability to produce surplus value is different from its indispensability for the realization of surplus value, as the realization entirely belongs to the sphere of circulation.

**Table 1**

*Classification of Consolidated Economic Activities in the United States into Productive and Non-productive Sectors*

| Industry | Include (1) or Exclude (0) |
|---|---|
| Farms | 1 |
| Forestry, fishing, and related activities | 1 |
| Oil and gas extraction | 1 |
| Mining, except oil and gas | 1 |
| Support activities for mining | 1 |
| Utilities | 1 |
| Construction | 1 |
| Wood products | 1 |
| Nonmetallic mineral products | 1 |
| Primary metals | 1 |
| Fabricated metal products | 1 |
| Machinery | 1 |
| Computer and electronic products | 1 |
| Electrical equipment, appliances, and components | 1 |
| Motor vehicles, bodies and trailers, and parts | 1 |
| Other transportation equipment | 1 |
| Furniture and related products | 1 |
| Miscellaneous manufacturing | 1 |
| Food and beverage and tobacco products | 1 |
| Textile mills and textile product mills | 1 |
| Apparel and leather and allied products | 1 |
| Paper products | 1 |
| Printing and related support activities | 1 |
| Petroleum and coal products | 1 |
| Chemical products | 1 |
| Plastics and rubber products | 1 |
| Wholesale trade | 0 |
| Retail trade | 0 |
| Transportation | 1 |
| Warehousing and storage | 1 |
| Information | 1 |
| Finance and insurance | 0 |
| Real estate | 0 |

| Rental and leasing services and lessors of intangible assets | 0 |
|---|---|
| Professional, scientific, and technical services | 1 |
| Management of companies and enterprises | 1 |
| Administrative and waste management services | 1 |
| Educational services | 1 |
| Health care and social assistance | 0 |
| Arts, entertainment, and recreation | 1 |
| Accommodation | 1 |
| Food services and drinking places | 1 |
| Other services, except government | 1 |
| Federal general government | 0 |
| Federal government enterprises | 0 |
| State and local general government | 0 |
| State and local government enterprises | 0 |

**Figure 2**

*Marxist ARoP Constructed with Sectors Included According to Established Criteria*

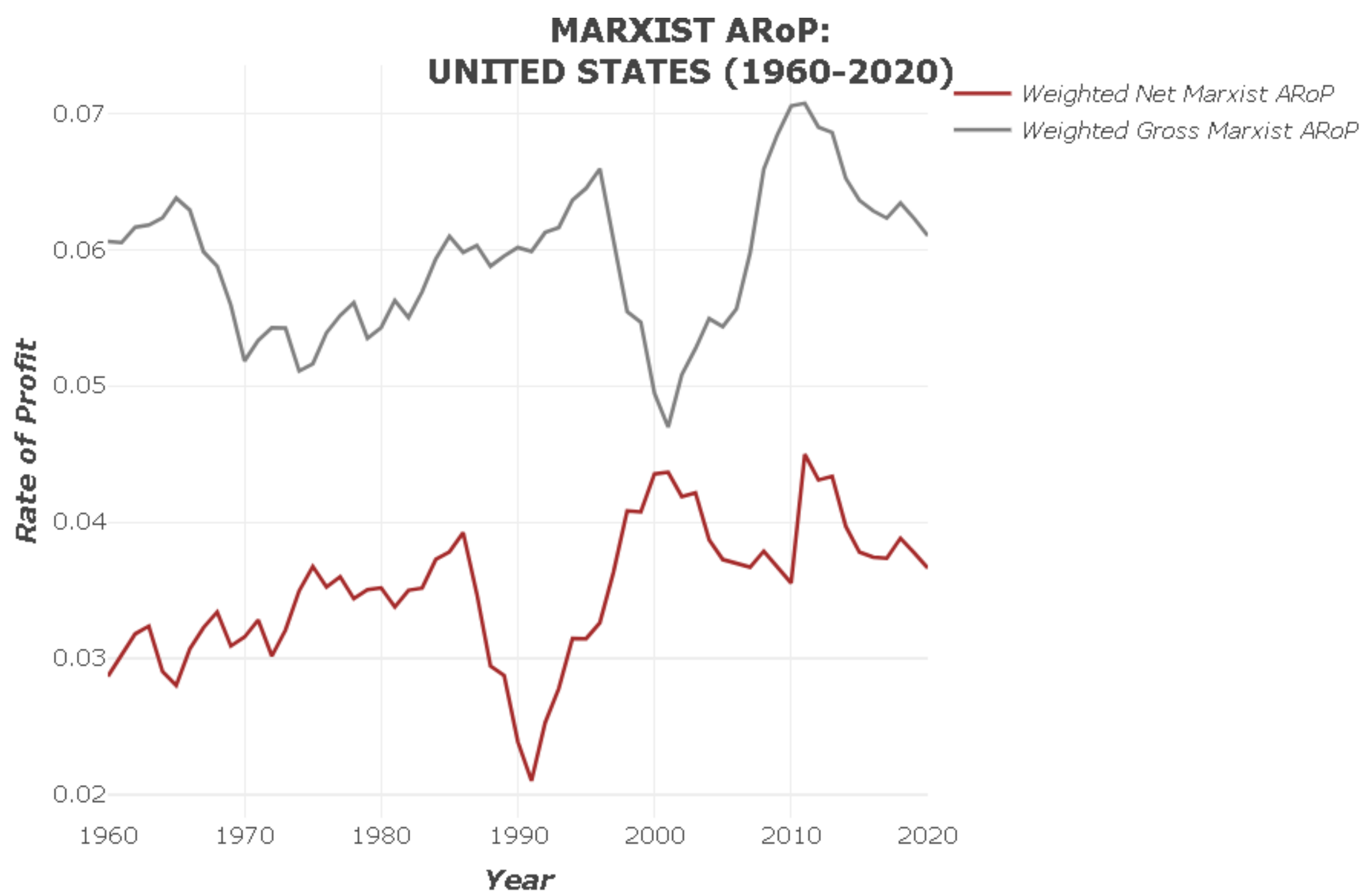

Most of the included and excluded sectors are easily understood considering the criteria already discussed. However, in some cases, the matter may not be as clear, and we will address them in the following section, referring to the US industrial classification system published in 2022.

The economic activities that may raise questions regarding their inclusion are "Utilities", "Warehousing and storage", "Professional, scientific, and technical services", "Educational services", "Administrative and waste management services", "Management of companies and enterprises", "Information" and "Other services, except government". We will now discuss these activities in detail. According to (Office of Management of Budget, 2022, p. 113), the utilities sector includes establishments engaged in providing utility services such as electric power (generation, transmission, and distribution), natural gas (distribution), steam supply (provision and distribution), water supply (treatment and distribution), and sewage treatment and disposal (collection, treatment, and waste elimination through sewer systems and wastewater treatment facilities). This sector excludes establishments primarily engaged in waste management services classified under Subsector 562[14]. Therefore, the inclusion of the utilities sector is justified based on the criteria previously established.

The inclusion of the sector of warehousing and storage, whose composition is defined in (Office of Management of Budget, 2022, p. 399), is justified because a wide variety of goods require different types of storage as part of maintaining their properties. This often involves not only the amortization of fixed capital but, more importantly, the technical handling of such equipment by the workforce (both complex and simple labor). Without this, the storage of goods would not be possible. Since the preservation of the properties that characterize the merchandise is an essential condition for the existence of the commodity as a product that will enter the sphere of circulation (note that this is still not part of the sphere of circulation), this economic activity must be included. Of course, there is a component in this sector that is exclusive to the sphere of circulation (related to distribution), and this component is a minority as its presence in this sector is defined as a possibility rather than a systematic fact[15].

Regarding professional, scientific, and technical services, it is stated in (Office of Management of Budget, 2022, p. 455) that these services require a high degree of expertise and training. They are primarily provided to businesses rather than for final consumption (although there is a minor component related to households), and they involve processes subcontracted by companies in predominantly productive sectors.

Regarding educational services, whose composition is defined in (Office of Management of Budget, 2022, p. 509) it should be noted that this economic activity may be one of the most controversial to include. It encompasses not only private education but also a public component and a non-profit component (related, when not public, to NGOs). However, excluding it would mean overlooking a fundamental component for the reproduction of a skilled workforce, which is particularly crucial in the context of a highly industrialized economy. The classification system mentioned does not specify the proportion that these two components represent in relation to the total sector.

[14] "Services of waste management and remediation" belongs to the sector "Administrative, Support, Waste Management and Remediation Services", which is the next sector under consideration.
[15] It is possible, as stated in the cited report, that "They may also provide a range of services, often referred to as logistics services, related to the distribution of gods".

The inclusion of waste man agement and administration services, whose composition is indicated in (Office of Management of Budget, 2022, p. 485), is justified because it encompasses companies that are primarily contracted by other businesses[16] (as part of subcontracting within the service sector in general) for waste management (cleaning and disposal), office administration, and administrative services related to document preparation [17], hiring and placement of personnel[18], as well as the preparation of documents related to productive sectors. All these components, if their services are provided to companies in productive sectors directly involved in value creation or those mentioned in footnote 41, should be included. It is evident that service activities such as solicitation, security, and surveillance (unless they are related to computer science and its derivatives[19]), and waste collection do not generate surplus value[20]. The proportion of activities within this economic sector that are linked to productive sectors is not specifically known, but it is publicly acknowledged[21] that most sectors in the economy analyzed are productive. Therefore, the decision to include them is made based on this understanding. Of course, this may vary when analyzing another country or even the same country in a different period.

Regarding the inclusion of the "Management of companies and enterprises" sector, the same argument as in footnote 16 should be made.

The inclusion of the information sector, whose composition is outlined in (Office of Management of Budget, 2022, p. 401), is justified because it produces and distributes information and cultural products, provides means to transmit or distribute these products, as well as data or communications, and processes data. Its main components include motion picture, sound recording, publishing (including software publishing), content and broadcasting providers, information technology infrastructure providers, data processing, web hosting (and related services), web search portals, libraries, archives, and other information services. In accordance with the criteria, it should be included.

---

[16] However, as stated, there is a minority component (presented as a possibility) related to household consumption.
[17] In the context of productive sectors, these services generate efficiency increases in related processes, thereby increasing the productive forces of labor and maximizing the rate of surplus value or exploitation of labor. In essence, these sub-processes aim to optimize the primary production process to maximize the ARoP.
[18] The same rationale as the previous footnote applies to this category.
[19] "This industry group comprises establishments primarily engaged in one of the following: (1) investigation, guard, and armored car services; (2) selling security systems, such as burglar and fire alarms and locking devices, along with installation, repair, or monitoring services; or (3) remote monitoring of electronic security alarm systems." (Office of Management of Budget, 2022, p. 497), indicating a significant component related to the field of Computer Science, specifically related to Information Technology.
[20] "This industry comprises establishments primarily engaged in collecting paymen ts for claims and remitting payments collected to their clients." (Office of Management of Budget, 2022, p. 492).
[21] Through input-output matrices.

The remaining services, excluding government, whose content is detailed in (Office of Management of Budget, 2022, p. 565), are included because they have an important component of equipment and machinery repair, funeral services[22], and defense (related to the defense industry). Thus, being consistent with the criteria outlined, despite containing sub-activities that are not value-generating, this sector should be included due to the relevance of its components to the reproduction of capital and the workforce.

Finally, the statistical results obtained demonstrate the consistency of the criteria constructed from Marxist theory in terms of the long-term trend behavior of the variable that this theory establishes as the fundamental variable of the capitalist political economy system.

### III.II. SEASONALITY ANALYSIS

If a spectral frequency analysis by a convolution smoothed periodogram via fast Fourier transform, as stated in (Kolmogórov & Fomin, 1978, p. 468), is performed in RStudio using the 'periodogram' syntax from the 'TSA' library, for sub-periods of 9 years[23], the following results are obtained.

**Table 2**

*Convolution Smoothed Periodogram via Fast Fourier Transform*

| Period | Primary Seasonality | Secondary Seasonality |
|---|---|---|
| 1960-1969 | 10 | 5 |
| 1969-1978 | 10 | 5 |
| 1978-1987 | 10 | 5 |
| 1987-1996 | 10 | 5 |
| 1996-2005 | 10 | 5 |
| 2005-2014 | 10 | 5 |
| 2014-2020 | 8 | 4 |

[22] Whether this labor is productive or unproductive, it is part of the costs of the workforce.
[23] Besides the mentioned economic reasons, statistical ones also played a role in the decision. The analysis revealed a primary seasonality of around 21 and a secondary seasonality of 16. To balance sample sizes and account for the overall small dataset, a sub-sample size of 9 was chosen, offering adequate sample size for pattern observation.

As can be observed, despite the last sub-period being 20% shorter than the other periods[24], a relatively constant variation in seasonality is observed, and in a strict sense, it does not exhibit an upward trend. Therefore, the use of additive filters is statistically justified.

The results show that the seasonality trend can be considered stable over time. As noted by (Polikar, 2003, p. 9), Fourier transformations, in general, can be used for non-stationary signals if we are only interested in knowing which spectral components exist in the signal but not interested in knowing where they occur, which is due to the assumption of stationarity of the signal made by the Fourier series and the Heisenberg's Uncertainty Principle (Gençay, Selçuk & Whitcher, 2002, p. 99).

But the latter does not imply that relevant information on the seasonality of the time series cannot be obtained, but rather that the performance of the instrument will be poor in terms of the quality of time resolution, which does not necessarily affect the exhibited seasonality trend and those kinds of transformations were applied not to see the time trend of the signal but rather the seasonality of the signal.

To address, or at least mitigate, the stationarity problem, one can use a Short Time Fourier Transform (STFT). As pointed out by (Gade & Gram-Hansen, 1996, p. 3), "The idea of the Short-time Fourier Transform, STFT, is to split a non-stationary signal into fractions within which stationary assumptions apply and to carry out a Fourier transform (FFT/DFT) on each of these fraction".

If a spectral frequency analysis via the STFT is performed in RStudio using the 'stft' syntax from the 'GENEAread' library, the following results are obtained.

**Figure 3**

*Short Time Fourier Transform Analysis Results*

[24]It is not implausible to assume that if the length of the last subperiod had been the same, the primary seasonality would be 5, and the secondary seasonality would be 10, as it happened in the second subperiod.

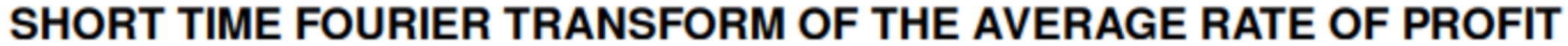

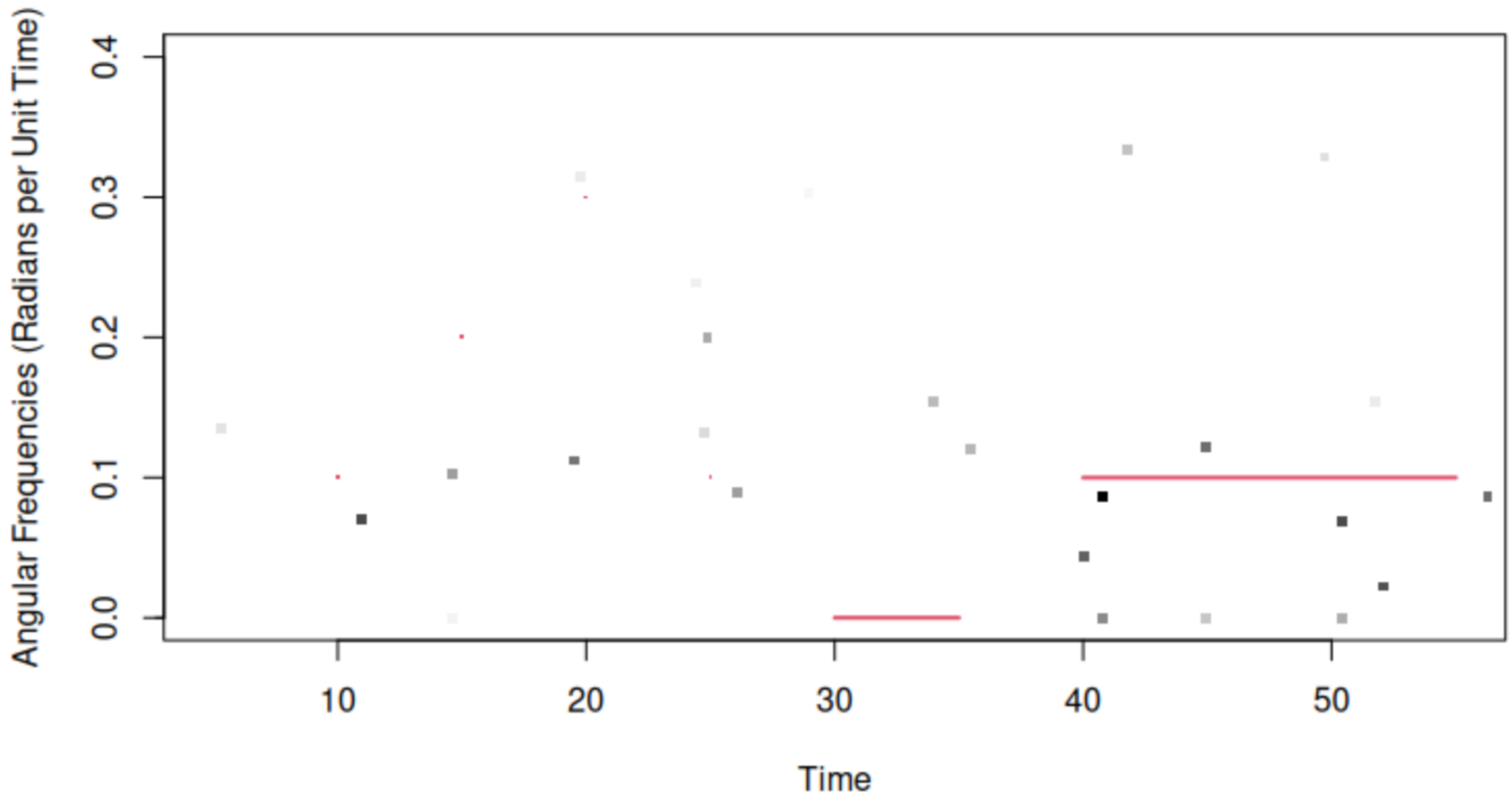

As can be observed, the seasonality trend can be considered stable over time. Together, all these results justify the use of additive filters.

## III.III. ECONOMETRIC ANALYSIS

### III.III.I. Trend of the Average Rate of Profit Under Our Proposed Criteria

#### III.III.I.I. Methodology for Calculating the Net Average Rate of Profit

Based on the data obtained from the Bureau of Economic Analysis (BEA), the ARoP was estimated as the weighted average[25] of sectoral profit rates. The sectoral profit rates were obtained by dividing the sum of sectoral surpluses by the total capital invested[26] in each sector.

Since disaggregated data for capital stock and intermediate consumption are not available, an assumption had to be made for their disaggregation. Regarding the capital stock, we assumed that capitalists participate in it in the same proportions as they participate in the total capital, or in words of (Ochoa, 1989, p. 427), "We then assumed further that the composition of the capital stock for each industry, in terms of the 71-commodity structure we are using, changes slowly over time. It follows that there is a straight- forward relation between the composition of gross investment—which is given for 1963,1967, and 1972—and the composition of the capital stocks." The same assumption was made regarding the proportion of intermediate consumption allocated to each productive sector.

---

[25] The weighting factor used was the proportion in which each sector participates in the total productive economy. See (Gómez Julián, 2023).
[26] This total sectoral capital consists of both constant and variable capital of the respective sector. Constant capital refers to the property, plant, and equipment in terms of stock, which should not be confused with its flow. Variable capital represents the total labor remunerations.

This means that the weights in the calculation of the overall average profit rate are based on the total capital of the productive sectors (not the entire economy), but the proportions used to estimate the volumes of capital stock and intermediate consumption are relative to the total economy. This is justified because the data for capital stock and intermediate consumption are aggregates for the entire economy.

### III.III.I.II. Less Asymmetric Daubechies Wavelet Transformation

If the syntax 'mra' from the 'waveslim' library is applied to estimate a DW with 8 vanishing moments (16 non-zero coefficients), abbreviated as $a8$, using a pyramid algorithm with a depth level[27] $J$ equal to 4, the following result is obtained.

**Figure 4**

*Marxist Net ARoP Trend (One-Year Lag)*

[27] The depth level (the number of iterations of the pyramid algorithm) has a maximum of $J = log_2 N$ (Gençay, Selçuk & Whitcher, 2002, pp. 135-136), as also indicated on page 51 of the library's manual. The value of $J$ is known as the depth of the decomposition. Higher depth levels (a higher value of $J$) result in a more detailed decomposition of frequencies in the time series, allowing for the identification of longer time cycles. The specific periodicity will vary depending on the particular dataset. For the sample size used in this research, the maximum of $J$ is 4.

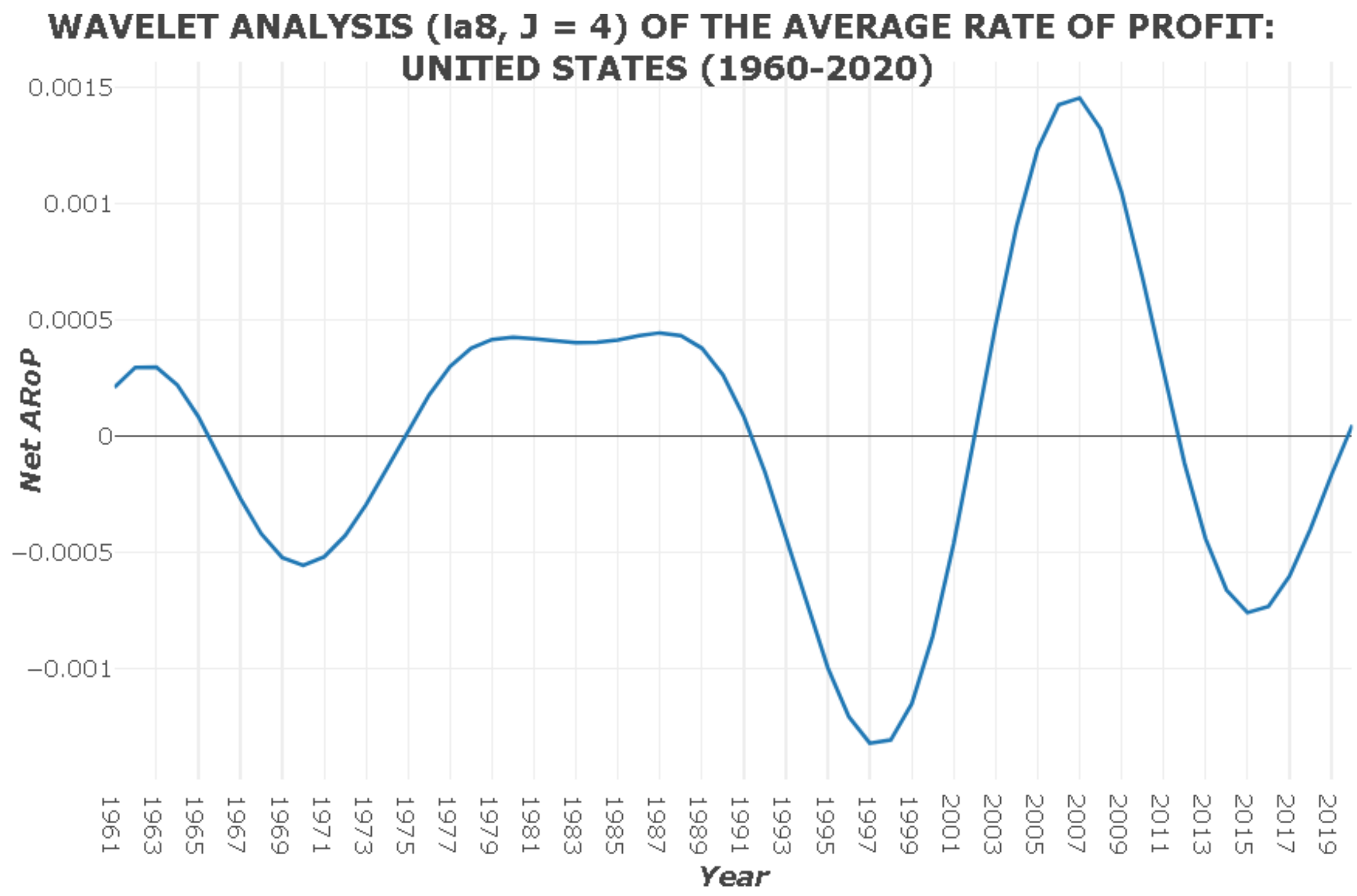

If the data is expressed in terms of levels, that is, by accumulating the results at different depth levels, the following result is obtained.

**Figure 5**

*Marxist ARoP Level-based Trend*

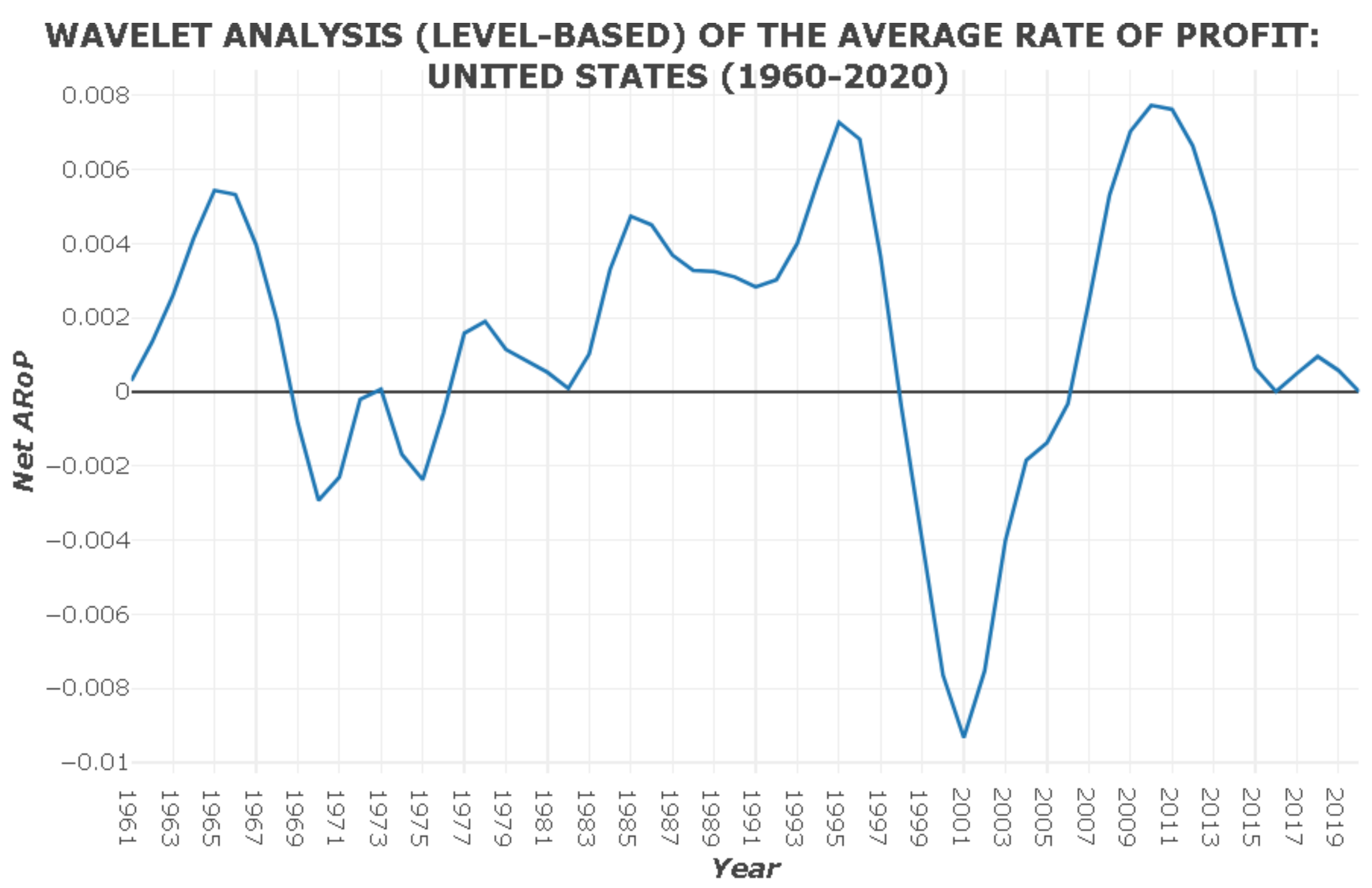

### III.III.I.III. Empirical Mode Decomposition Transformation

Applying the 'emd' syntax from the 'EMD' library, the following result is obtained.

**Figure 6**

*Trend using the Empirical Mode Decomposition of the Marxist Net ARoP*

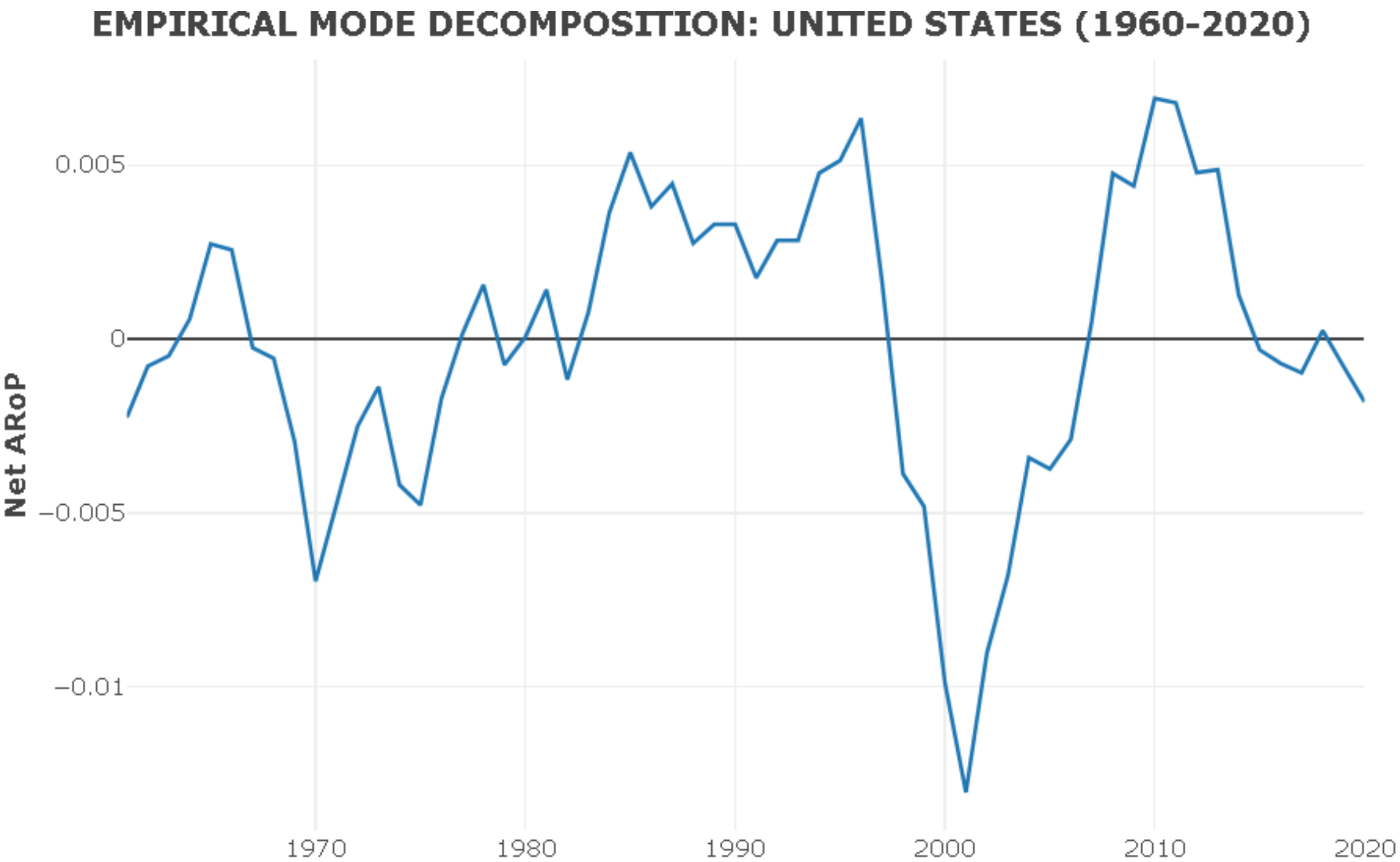

### III.III.I.IV. Transformation using Hodrick-Prescott Filter Embedded in Unobservable Variables Model

Since this filter uses the Gibbs sampler, one can choose the average of the sequence of iteration results (Chan et al., 2019, p. 377) or the last value of the simulation (Casella, 1992, p. 168). We chose the average of the sequence iteration because it is the posterior mean of the distribution of simulations.

**Figure 7**

*Average Trend of the Marxist Net ARoP*

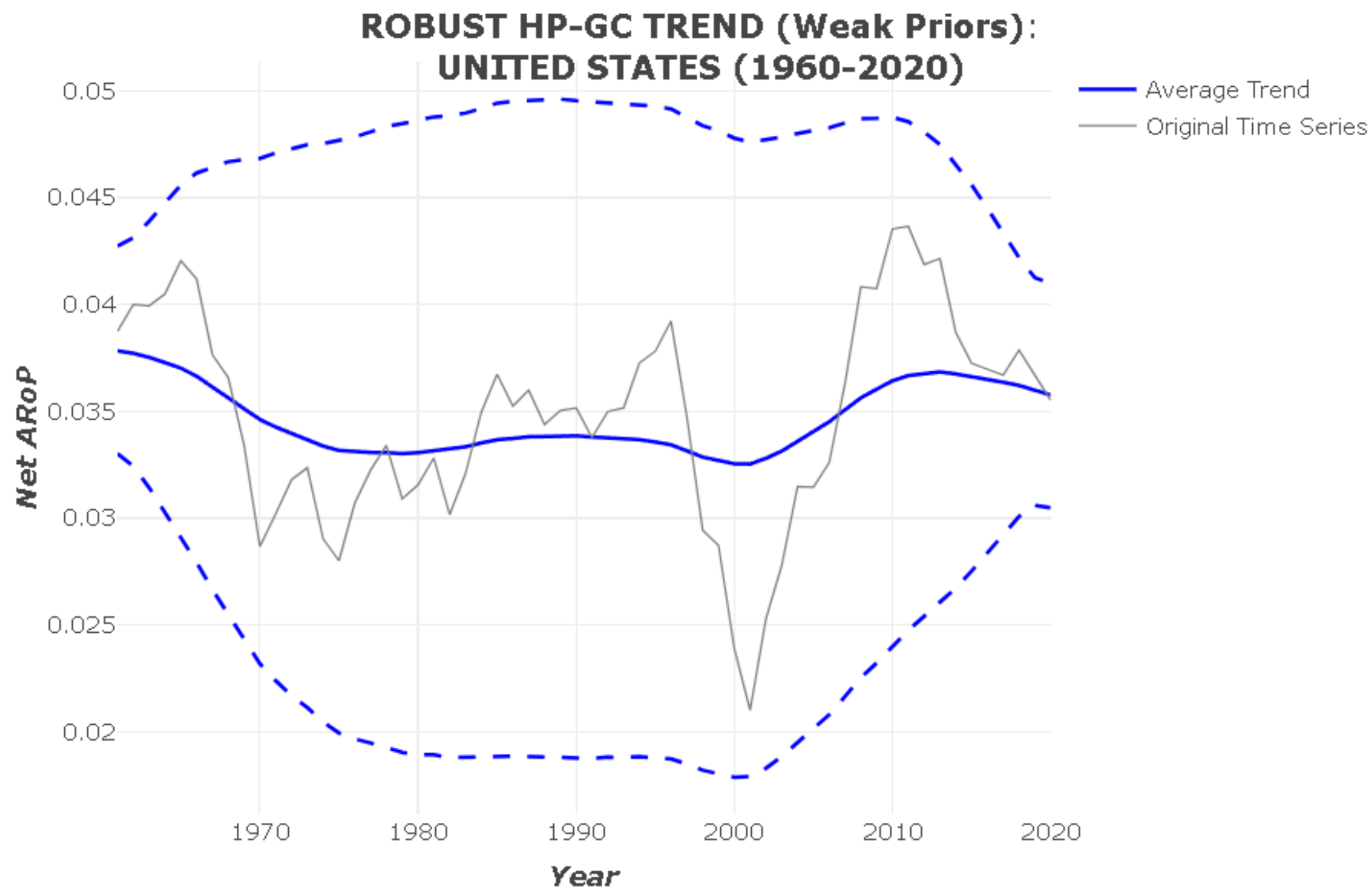

### III.III.II. Unit Root Tests

Below are presented the statistical results of the three variations of the Augmented Dickey-Fuller test (ADF) (Said & Dickey, 1984, pp. 600-601), with 'ur.df' syntax, the six variations of the Elliott, Rothenberg, and Stock test (ERS) (Elliott, Rothenberg & Stock, 1996, pp. 815-826) with 'ur.ers' syntax, the four variations of the Kwiatkowski test (KPSS) (Kwiatkowski, et al. 1992, pp. 162-164) with 'ur.kpss' syntax, and the four variations of the Phillips-Perron test (PP) (Perron & Ng 1996, pp. 436-441) with 'ur.pp' syntax.

**Table 3**

*Unit Root Tests for Marxist ARoP (Net and Weighted): Is the Null Hypothesis Rejected?*

| TEST TYPE | TEST VARIATION | SIGNIFICANCE LEVEL | | |
|---|---|---|---|---|
| | | ***0.01*** | ***0.05*** | ***0.1*** |
| *ADF* | No Drift or Deterministic Trend | No | No | No |
| | Drift but No Deterministic Trend | No | No | No |
| | Drift and Deterministic Trend | No | No | No |
| *ERS (DF-GLS[28])* | Constant Mean | No | Yes | Yes |
| | Linear Trend | No | No | No |
| | Constant Mean and Linear Trend | No | Yes | Yes |

[28] Dickey–Fuller with Generalized Least Squares Detrending.

| | | | | |
|---|---|---|---|---|
| | Constant Mean Influenced by Residual Autocorrelation | No | Yes | Yes |
| *ERS (P-test[29])* | Linear Trend Influenced by Residual Autocorrelation | No | No | No |
| | Constant Mean and Linear Trend Influenced by Residual Autocorrelation | No | Yes | Yes |
| *KPSS* | Short Lags around a Random Walk | No | No | No |
| | Short Lags around a Deterministic Trend | No | No | Yes |
| | Long Lags around a Random Walk | No | No | No |
| | Long Lags around a Deterministic Trend | No | No | No |
| *PP* | Short Largs with $Z(\alpha)$ | No | No | No |
| | Short Lags with $Z(t)$ | No | No | No |
| | Long Lags with $Z(\alpha)$ | No | No | No |
| | Long Lags with $Z(\alpha)$ | No | No | No |

As noted by Davidson & MacKinnon (2004, p. 613), there is substantial evidence that Phillips-Perron tests perform less well in finite samples than ADF. Consistently, Elliott, Rothenberg & Stock (1996, 830) show that their DF-GLS procedure has superior finite-sample power relative to ADF.

In our battery of unit-root tests, PP ($Z_\alpha$, $Z_\tau$) fails to reject a unit root, ERS DF-GLS rejects under a constant (and the ERS P-test rejects under constant and constant + trend), while KPSS mostly fails to reject stationarity—especially around a deterministic trend. Taken together with the preceding wavelet, EMD, and Bayesian UCM-HP analyses—which consistently isolate a dominant low-frequency component—these findings are most consistent with ARoP exhibiting a deterministic (potentially non-linear) trend. We therefore refrain from imposing any specific parametric trend and interpret the unit-root results in light of the nonparametric trend revealed by those filters and the definition of stationarity (Cryer & Chan, 2008, pp. 16-17).

**III.III.III. Trend of the Average Rate of Profit Under Different Statistical Criteria for Sector Selection**

**III.III.III.I. Principal Component Analysis (PCA) Via Singular Value Decomposition (SVD)**

Instead of Johnson–Lindenstrauss (JL) random-projection embeddings, which is the alternative par excellence to approximate isometries (Matoušek, 2013, pp. 26–29), we perform deterministic dimensionality reduction using principal component analysis (PCA) via singular value decomposition (SVD), yielding orthogonal directions that maximize variance and provide uncorrelated principal components for inference (Jolliffe & Cadima, 2016, pp. 1-3, 8).

[29] Point-Optimal test.

The PCA can be executed using the 'PCA' syntax from the 'FactoMineR' library. Then, the scree plot can be constructed using the 'fviz_screeplot' syntax from the 'factoextra' library. Following the rule of including the principal components (PC) whose eigenvalue is greater than unity and that together represent between 70% and 90% of the variance (Everitt & Hothorn, 2011, p. 71). Furthermore, the linear independence of the PC can be analyzed using the 'corrplot' syntax of the 'corrplot' library.

**Figure 9**

*Principal Components Screen Plot (All Economic Activities)*

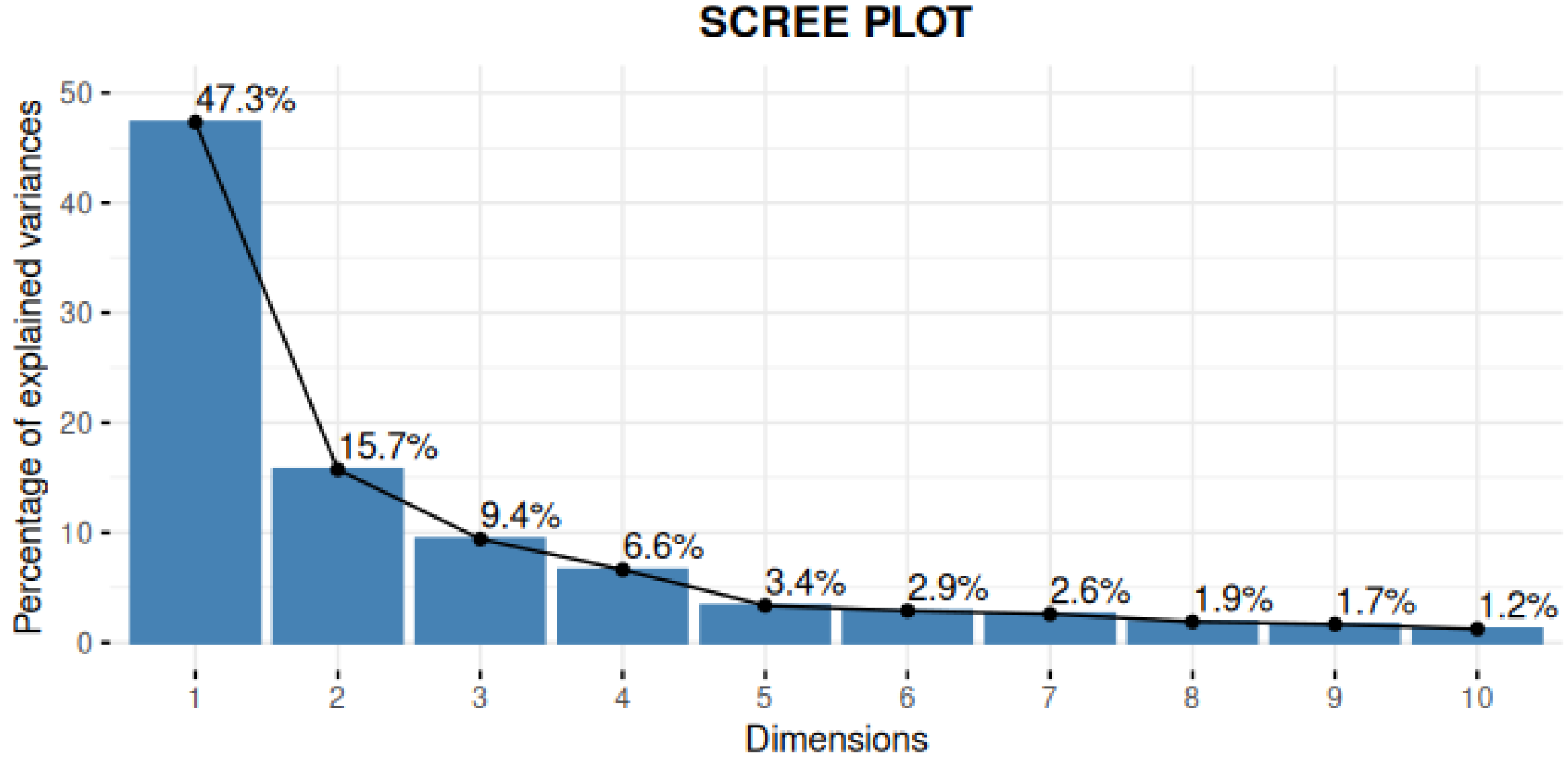

**Figure 10**

*Linear Independence Analysis of PC*

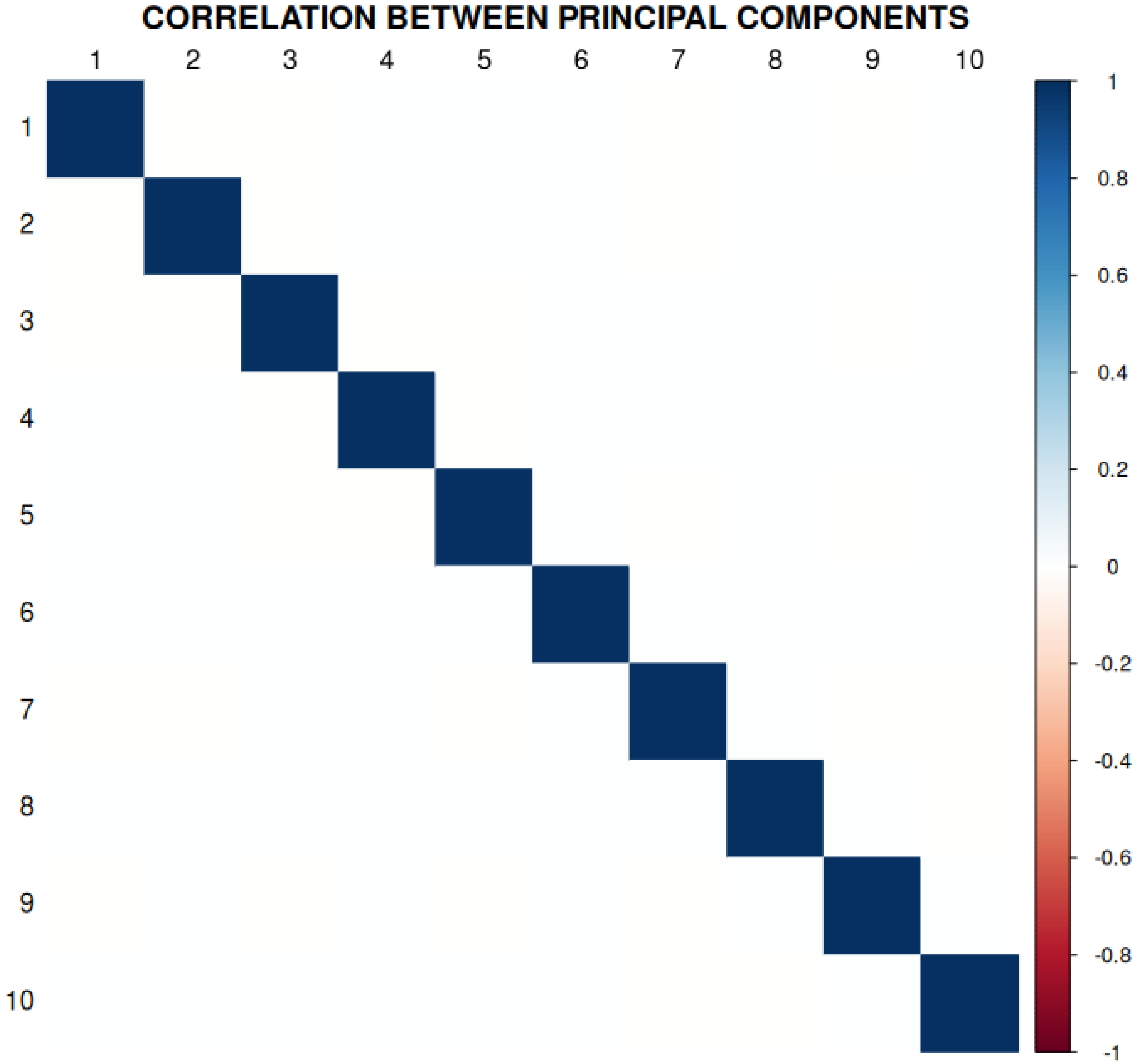

In geometric terms, the relevance of 'large' SV lies in the fact that the ellipsoid's axis will be significantly longer than the corresponding axis of a sphere. However, there is no standard criterion to determine what is 'large' or 'small'.

Given the multivariate nature of the dataset and the need to identify which productive sectors act as the true structural drivers of economic variance, arbitrary selection criteria—such as the simple arithmetic mean—were discarded in favor of a rigorous geometric and probabilistic approach. Geometrically, the relevance of a "major" sectoral contribution lies in its capacity to extend the axis of the variance ellipsoid significantly beyond what would be expected under a uniform or spherical distribution. However, given the absence of a priori standard to determine the threshold between a significant and a residual contribution, we proceeded to empirically model the probability distribution function of the sectoral contributions for each of the seven retained Principal Components (PCs).

To ensure the robustness of this modeling, an exhaustive statistical competition procedure was implemented using the `fitdistrplus` library in R. The percentage contributions of the 47 economic sectors were subjected to fitting against eight candidate theoretical distributions: Normal, Log-Normal, Gamma, Weibull, Cauchy, Logistic, Uniform, and Student's t. Unlike standard approaches relying on a single estimation method, this analysis contrasted parameter stability using four simultaneous estimation methods: Maximum Likelihood Estimation (MLE), Maximum Goodness-of-Fit Estimation (MGE) optimizing the Cramer-von Mises distance, Method of Moments (MME), and Maximum Spacing Estimation (MSE).

The selection of the optimal model for each dimension was performed by minimizing the Bayesian Information Criterion (BIC), thereby prioritizing model parsimony and penalizing overfitting. The results of this distributional analysis revealed a clear structural pattern: while the first principal component (PC1), associated with the general size of the economy, showed a more homogeneous distribution (Uniform), the subsequent components (PC2 to PC7) exhibited markedly asymmetric behaviors, fitting better to heavy-tailed distributions such as Gamma and Weibull. This finding confirms that the structural variance of the United States economy is not distributed evenly; rather, it is highly concentrated within a small number of dominant sectors, while most activity branches contribute only marginal statistical noise.

Consistent with this evidence of positive skewness, a strict cutoff criterion was established based on the location of the sectors in the upper tail of the empirical distribution. Specifically, only those sectors whose contribution fell within the top decile (or Top 10%) of the fitted distribution for each component were retained. This procedure allowed for the isolation of the structural signal from the noise, reducing the dimensionality of the analysis from 47 original variables to a consolidated core of 26 strategic sectors.

The result, visualized via an intensity Heatmap, demonstrates a clear functional specialization within the economy. Distinct sectoral clusters were identified: an axis of corporate and financial services dominating PC1; a logistics-industrial chain characterizing PC2; an extractive natural resources block defining PC4; and a critical social welfare component (PC7) driven predominantly by educational services and public administration. This methodological filter excluded 21 sectors (including agriculture and construction) whose joint variance proved negligible in explaining the dynamics of the selected principal components. This validates that the covariance structure of the data—and consequently, the economic interpretation of the PCs—remains invariant despite the exclusion of these lower-hierarchy variables.

**Figure 11**

*Heatmap of United States Economy*

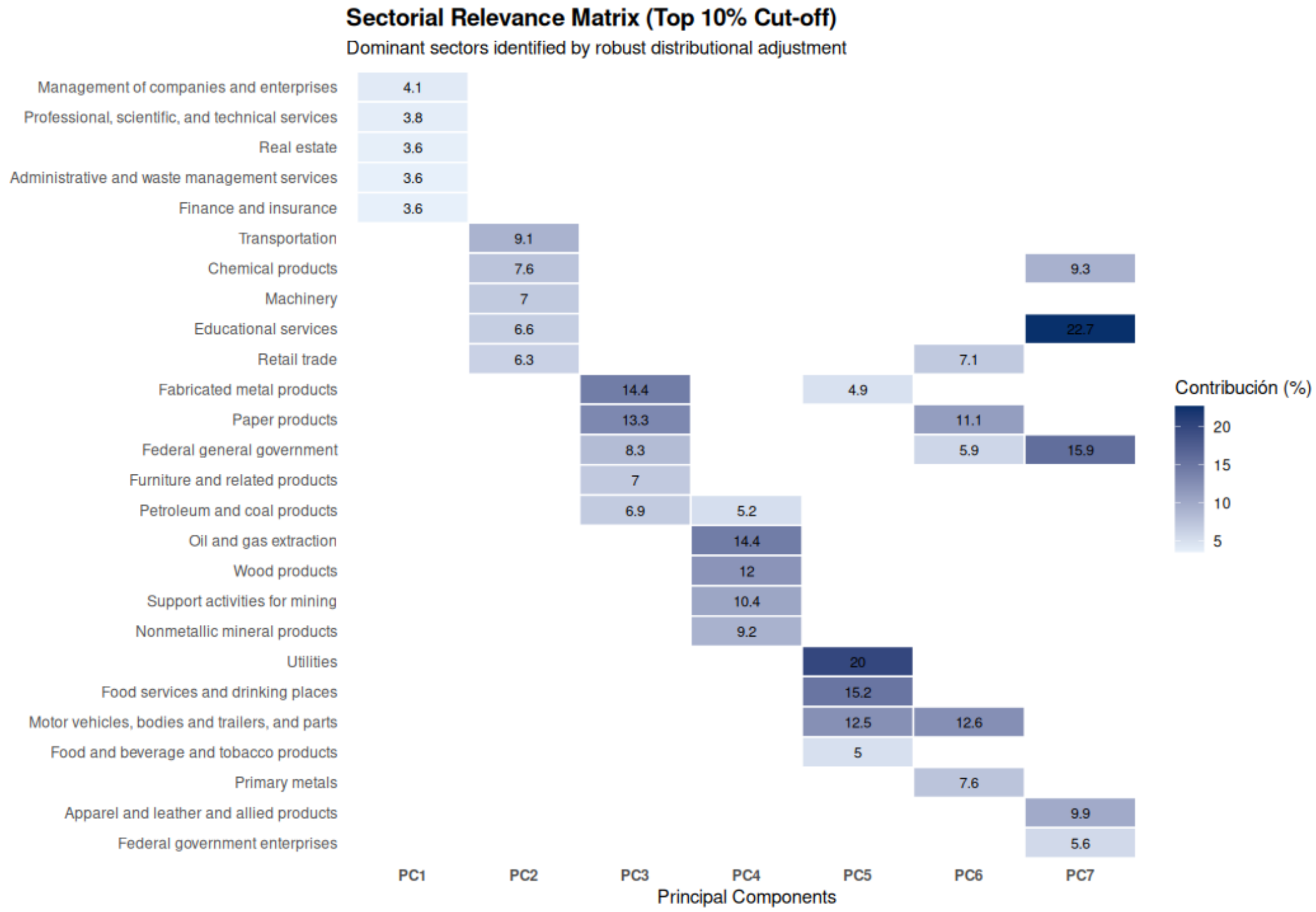

To robustify the exclusion decision and empirically verify the hypothesis of structural invariance, the set of 21 discarded variables was subjected to a post hoc validation test (see Table 4). This procedure consisted of recalculating the *absolute inertia load* of each excluded sector, transforming their relative percentage contributions back into eigenvalue units. Following the logic of the standard Kaiser criterion, for a variable or subset of variables to be considered an independent latent driver capable of distorting the covariance structure, its equivalent eigenvalue must exceed unity ($\lambda > 1$).Analysis of the excluded sectors reveals that none of the 21 discarded branches of activity reaches this critical threshold in any of the first five principal dimensions. The equivalent eigenvalues consistently fall below unity, with a maximum of 0.77 (observed in Health care for the first dimension) and weighted averages that barely reach the 0.25 to 0.40 range. This mathematically demonstrates that these sectors lack sufficient statistical "energy" to constitute a latent dimension on their own. Consequently, their exclusion does not imply a loss of structural information, but rather an effective removal of redundancies, confirming that these sectors act as passive recipients of aggregate economic dynamics rather than as determinant drivers.

**Table 4**

*Eigenvalues of Excluded Sectors in Relevant PC*

| SECTOR | DIM 1 | DIM 2 | DIM 3 | DIM 4 | DIM 5 | Simple Mean | Weighted Mean |
|---|---|---|---|---|---|---|---|
| Farms | 0,5 | 0,07 | 0,19 | 0,04 | 0 | 0,16 | 0,27 |
| Forestry, fishing, and related activities | 0,51 | 0,01 | 0,17 | 0,01 | 0,01 | 0,14 | 0,26 |
| Mining, except oil and gas | 0,71 | 0,09 | 0 | 0,1 | 0 | 0,18 | 0,36 |
| Construction | 0,63 | 0,01 | 0,04 | 0,11 | 0,07 | 0,17 | 0,31 |
| Computer and electronic products | 0,3 | 0,42 | 0,05 | 0,14 | 0 | 0,18 | 0,22 |
| Electrical equipment, appliances, and components | 0,61 | 0,07 | 0,03 | 0,03 | 0,04 | 0,16 | 0,3 |
| Other transportation equipment | 0,73 | 0,2 | 0,01 | 0 | 0 | 0,19 | 0,38 |
| Miscellaneous manufacturing | 0,46 | 0,1 | 0 | 0,15 | 0,04 | 0,15 | 0,25 |
| Textile mills and textile product mills | 0,57 | 0,01 | 0,06 | 0,07 | 0,02 | 0,15 | 0,28 |
| Printing and related support activities | 0,55 | 0,27 | 0,08 | 0 | 0,05 | 0,19 | 0,31 |
| Plastics and rubber products | 0,72 | 0,16 | 0,06 | 0,01 | 0 | 0,19 | 0,37 |
| Wholesale trade | 0,52 | 0,2 | 0,02 | 0,08 | 0,02 | 0,17 | 0,29 |
| Warehousing and storage | 0,71 | 0,1 | 0,05 | 0 | 0,04 | 0,18 | 0,36 |
| Information | 0,6 | 0,29 | 0,04 | 0 | 0,01 | 0,19 | 0,33 |
| Rental and leasing services and lessors of intangible assets | 0,7 | 0 | 0,01 | 0 | 0,02 | 0,15 | 0,33 |
| Health care and social assistance | 0,77 | 0,2 | 0 | 0,02 | 0 | 0,2 | 0,4 |
| Arts, entertainment, and recreation | 0,73 | 0,11 | 0 | 0,03 | 0,02 | 0,18 | 0,37 |
| Accommodation | 0,62 | 0,18 | 0 | 0,02 | 0,03 | 0,17 | 0,33 |
| Other services, except government | 0,71 | 0,13 | 0,02 | 0,01 | 0 | 0,17 | 0,36 |
| State and local general government | 0,56 | 0,38 | 0,01 | 0,02 | 0 | 0,19 | 0,33 |
| State and local government enterprises | 0,74 | 0,11 | 0,06 | 0 | 0 | 0,18 | 0,37 |

**III.III.III.II. Regularized Horseshoe Regression**

The name 'Horseshoe' is not a poetic metaphor but a literal description of the geometry governing its shrinkage coefficient ($\kappa$). In technical terms, this method utilizes a hierarchy of half-Cauchy distributions for local and global scale parameters, inducing a probability density for the shrinkage factor $\kappa \in (0,1)$ in a symmetric 'U' shape—resembling a vertical horseshoe—featuring asymptotic mass accumulations (infinite spikes) at the extremes $\kappa = 0$ and $\kappa = 1$. This specific topology en-dows the model with its 'global-local shrinkage' property, enabling it to aggressively discriminate noise by crushing irrelevant coefficients toward 0 with immense force, while simultaneously protecting strong signals with near-total robustness (leaving them intact near 1), thus overcoming the uniform shrinkage bias typical of methods such as LASSO.

The Regularized Horseshoe regression model was specified by setting the Total Gross Operating Surplus as the dependent variable, predicted by the Total Variable Capital across all sectors. This formulation was strategically chosen to empirically ground the variable selection process in the labor theory of value, positing variable capital as the active structural determinant of surplus value generation, while leveraging the model's shrinkage properties to rigorously identify the most relevant sectoral drivers amidst high multicollinearity.

The structure of economic interdependence inherent in an Input-Output Matrix generates a structural multicollinearity that invalidates the use of conventional variable selection algorithms. Methods such as backward elimination or frequentist LASSO were discarded due to their arbitrary handling of correlation: when faced with two productive sectors that strongly covary, these algorithms tend to select one and suppress the other based on minor stochastic fluctuations. This would introduce false negative biases incompatible with value theory. Likewise, the use of Bayesian Generalized Linear Models (GLM) with non-informative priors was dismissed, given that high collinearity under these conditions results in diffuse and indistinguishable posterior distributions, preventing a robust causal identification of surplus-value generating sectors.

To resolve this high-dimensional challenge ($p \approx n$) while preserving inferential stability, we opted for a Bayesian regularization approach using the *Regularized Horseshoe* prior. Although the *Spike-and-Slab* prior theoretically offers the purest variable selection via point masses at zero, its computational implementation in Hamiltonian Monte Carlo (HMC) engines proves unstable when dealing with highly correlated predictors, generating convergence issues and poor chain mixing. The Regularized Horseshoe was selected as the superior alternative due to its global-local shrinkage property: it utilizes continuous distribution that severely compresses noise toward zero (simulating sparsity) but maintains heavy tails that allow strong signals to escape regularization without bias, thus facilitating the algorithm's navigation over the posterior probability surface.

Horseshoe regression is structurally defined using a specific hierarchy of prior distributions for the coefficients. Given that it relies on the specification of priors and updating via likelihood to obtain a posterior, it is intrinsically a Bayesian method. The model implementation was performed on log-transformed data to interpret coefficients as elasticities and to stabilize variance. The configuration of the Horseshoe hyperparameters was designed to balance parsimony with signal detection: we established an *a priori* expectation of sparsity ($p_0$) of 16 sectors—approximately one-third of the total—combined with a "slab" scale of 2 and 4 degrees of freedom. This heavy-tailed specification (Student-t) in the *slab* component was deliberate to ensure that the model, while skeptical of noise, remained flexible enough to accommodate coefficients of large magnitude without imposing excessive shrinkage, thereby overcoming the limitations of more restrictive priors such as the Laplace (Bayesian LASSO).

Given the geometric complexity of the posterior distribution derived from multicollinearity, estimation was executed using the *No-U-Turn Sampler* (NUTS) algorithm in Stan, reinforced with QR decomposition. (Note: Given the severe multicollinearity of the Input-Output Matrix, the design matrix $X$ exhibits a pathologically high condition number ($\kappa(X) \gg 30$), which distorts the geometry of the posterior distribution, turning it into a narrow, elongated "canyon" that hinders efficient exploration by the Hamiltonian algorithm. To mitigate this numerical conditioning problem, we implemented a reparameterization based on QR Decomposition, which factorizes the predictor matrix such that $X = QR$, where $Q$ is an orthogonal matrix ($Q^T Q = I$) and $R$ is upper triangular. This transformation rotates the parameter space, allowing MCMC sampling to be performed on a vector of transformed coefficients $\tilde{\beta}$ associated with the orthogonal vectors of $Q$. Being orthogonal, these transformed predictors eliminate *a priori* correlation in the sampling process, inducing a more spherical and Euclidean posterior geometry that facilitates Markov chain convergence and mixing. Subsequently, the original coefficients of interest are deterministically recovered via the inverse relationship $\beta = R^{-1}\tilde{\beta}$, thus achieving numerical stability without altering the substantive inference of the model). We applied an extremely rigorous sampling strategy, configuring 4 chains with an extended *warmup* period of 4000 iterations and a target acceptance rate (adapt_delta) of 0.999. This robust configuration was necessary to avoid divergent transitions and ensure impeccable convergence diagnostics ($\hat{R} < 1.01$) before proceeding to the final phase, where the definitive selection of variables was not based on the raw magnitude of the coefficients, but on the predictive projection technique (with *projpred* function of the 'rstanarm' library), which seeks the minimal submodel capable of reproducing the predictive capacity of the regularized reference model.

According to the results of the run, no sector is statistically significant (which is theoretically and practically impossible). What occurred is evident: to control for multicollinearity, the model "crushed" every signal, leaving nothing behind after multiple controls. Thus, it produced multiple false positives regarding the noise, treating all multicollinearity as noise when, in fact, multicollinearity is an ontological characteristic of the US input-output system.

Although the Regularized Horseshoe is designed with heavy tails (Cauchy/Student-t) theoretically capable of letting strong signals "escape" global shrinkage, this only occurs if the likelihood of the data provides sufficient evidence to counteract the prior's pressure toward zero.

In our case, the fact that even with the Horseshoe all coefficients collapsed ($\kappa \approx 1$, selection of 1 variable) suggests empirically that there are no independent strong signals in the data, a view supported by other econometricians as discussed below. The model did not "fail" to let signals escape; rather, it correctly diagnosed that, once the global correlation structure is controlled, what remains is indistinguishable from noise. In practice, it behaved like a Spike-and-Slab where the probability mass concentrated almost entirely on the "Spike" (zero) because the "Slab" (the signal) lacked empirical support. This posits that predictive ranking is the only salvageable metric, given that the magnitude of individual effects is statistically evanescent.

Christopher Achen (Gujarati, 2009, p. 326) notes that:

> The only effect of multicollinearity has to do with the difficulty of obtaining estimated coefficients with small standard errors. However, the same problem arises when having a small number of observations or independent variables with small variances. (In fact, at the theoretical level, the concepts of multicollinearity, small number of observations, and small variances in independent variables are essential parts of the same problem). Therefore, the question "what should be done then with multicollinearity?" is similar to "what should be done if you don't have many observations?" There is no statistical answer to this.

In the same vein, Blanchard (Gujarati, 2009, p. 342) metaphorically suggests that "Multicollinearity is God's will, not a problem with OLS or with statistical technique in general."

The underlying logic of these statements is identical. In Achen's case, it signifies limits in statistical methodologies, while for Blanchard, it posits that Nature is intrinsically intensely correlated and that, faced with this fact, there is nothing to do but accept the prediction difficulties given the limitations of our measurement instruments. This makes taking the 85th or 90th percentile (TOP 15 or TOP 10) theoretically and statistically valid, since multicollinearity is the reality—that is, the ontology of the system: in an Input-Output Matrix, if the "Steel" sector grows, the "Automobiles" sector must grow. Treating that correlation as a statistical "error" is to ignore the material structure of production. As Blanchard says, it is "God's will" (or in Marxist terms, the law of value operating systemically).

We have 61 years of data to explain 47 sectors that move in unison. Achen is right: it is a problem of "lack of independent information". One could say that there is no mathematical trick that creates information where there is none.

However, although uncertainty prevents us from stating "V28 is definitely distinct from zero and V36 is not", the *projpred* algorithm calculates that V28 reduces predictive error faster than V36. That ordering is valid and valuable information, even if traditional "statistical significance" is diluted.

An alternative to "creating information where there is none" could be the execution of an empirical distribution adjustment from a maximum robustness Bayesian approach (implementing best practices to handle uncertainty, including MCMC with NUTS), synthetically extending the sample (to 5,000 or 10,000 observations), and then re-running the Horseshoe.

However, as we have already advanced, the correlation observed between sectors in an Input-Output Matrix does not constitute a "statistical problem" susceptible to technical correction, but the phenomenal manifestation of the law of value operating systemically; it is the expression of the material interdependence of the capitalist production process, where the covariance between steel and automobiles is not "noise", but structure. Consequently, the Regularized Horseshoe approach, by treating all correlation as redundancy subject to shrinkage, incurs an ontological error by confusing the constitutive structure of the object with the imprecision of the measuring instrument. The proposal to extend the sample through synthetic data generation does not resolve this ontological limitation, as fitting a distribution to the 61 historical years and sampling 10,000 synthetic observations merely replicates the original correlation structure without introducing independent variation among predictors. From the perspective of historical materialism, the data ($N = 61$) represent the real and finite history of capital accumulation; pretending to create 10,000 synthetic years constitutes an idealist exercise that fabricates a history that never occurred to force a mathematical conclusion, ignoring Blanchard's maxim that multicollinearity must be accepted as an intrinsic property of the nature of macroeconomic data.

Epistemically, following Achen, multicollinearity, reduced sample size, and low variance are manifestations of the same epistemic limit: the insufficiency of independent information to isolate causal effects. What the model requires is not more observations of the same joint distribution $P(V_{1,}V_{2,}\ldots,V_{47})$, but orthogonal variation between sectors—something ontologically impossible given that capitalism does not permit isolated production dynamics. Synthetic sampling operates under a circular logic analogous to photocopying a blurry text with the hope of gaining legibility; it violates the principle that it is impossible to extract more information than the data originally contains. In econometrics, "there is no free lunch": available information is limited to the 61 bits of independent history, and any attempt to artificially inflate $N$ constitutes an "Informational Bootstrap Fallacy". This leads to false methodological precision, artificially reducing standard errors and silencing the honest warning of uncertainty that the original model yields, producing results that are technically significant but epistemically empty.

Therefore, epistemic honesty demands abandoning the pretension of confirmatory statistical significance in an environment of $N = 61$ and $P = 47$, validating instead the predictive projection ranking (*projpred*) as the robust finding. Although credibility intervals cross zero due to collinearity, the ordering of variables according to their capacity for predictive error reduction (ELPD) constitutes real, non-random information. Since no statistical solution exists for the lack of independent variation, the correct strategy is to utilize Marxist theory as the substantive prior to interpret this exploratory ranking, accepting the historical limitations of the data as they are, rather than fabricating artificial certainty.

**Table 5**

*Ranking Position of Economic Activities by Their Prediction Power in Regularized Horseshoe Regression*

| Ranking Position | Economic Activity |
|---|---|
| 1 | Retail Trade |
| 2 | Textile Mills and Products |
| 3 | Fabricated Metal Products |
| 4 | Administrative and Waste Management Services |
| 5 | Miscellaneous Manufacturing |
| 6 | Construction |
| 7 | Educational Services |
| 8 | Electrical Equipment, Appliances, and Components |
| 9 | Nonmetallic Mineral Products |
| 10 | Support Activities for Mining |
| 11 | Printing and Related Support Activities |
| 12 | Primary Metals |
| 13 | Food Services and Drinking Places |
| 14 | State and Local General Government |
| 15 | Transportation |

Once the productive core was delimited via variable selection using the Regularized Horseshoe and predictive projection, the aggregate Net Average Rate of Profit was calculated exclusively for this subset of 15 strategic sectors. To isolate the secular trend from the cyclical fluctuations and high-frequency noise inherent in annual data, the resulting time series was subjected to three signal extraction algorithms of distinct mathematical natures: Empirical Mode Decomposition, Daubechies Wavelet filtering, and the Bayesian Hodrick-Prescott filter embedded within an unobserved components state space.

A comparative analysis of the extracted signals reveals a robust empirical consensus: regardless of the decomposition technique employed, the trajectory of the profit rate for the US productive core exhibits an unequivocally declining secular trend over the 1960–2020 period. However, the morphology of this decline presents structural nuances depending on the algorithm; while EMD and Wavelets capture low-frequency oscillations associated with long waves or shifts in accumulation regimes, the trend extracted via the HP-GC filter manifests as the most rigid and monotonic trajectory of the set.

This accentuation of the trend in the case of the HP-GC filter is not a random artifact but a direct consequence of its state-space specification, which mathematically penalizes the variance of the trend component’s second difference (i.e., its acceleration). By imposing a 'global memory' structure on the series, the algorithm forces a smoothing process that connects the entire history of the data, thereby restricting the model's capacity to adapt to abrupt regime changes or local volatility. Consequently, this configuration tends to linearize the signal extraction process, resulting in a trend curve that, while sacrificing reactivity to conjunctural shocks, more clearly crystallizes the underlying historical direction of profitability. This confirms, for the defined economic activities, the Law of the Tendency of the Rate of Profit to Fall not as a cyclical phenomenon, but as a long-term gravitational force operating upon the economy's productive core.

### III.III.III.III. Dyamic Factor Auto-Regressive Model

An estimation of a dynamic factor auto-regressive model, as defined by (Krantz & Bagdziunas 2023, pp. 7-9), can be carried out through 'DFM' syntax from library 'dfms'. By $PC_{p3} = V(k) = \left(k, \hat{F}^k\right) + k\hat{\sigma}^2 \left(\frac{lnC_{NT}^2}{C_{NT}^2}\right)$ criterion, where $PC$ are the PC, $N$ is the cross section dimension (47 in our case), $T$ is the time dimension (60), $F$ es common factors vector ($\hat{F}$ its estimation), $k < min[N, T]$ is an arbitrary seed number for the simulation, $C_{NT}$ is $min\left(\sqrt{N}, \sqrt{T}\right)$, $\sigma^2$ is the consistent estimation of $(NT)^{-1} \sum_{i=1}^{N} \sum_{t=1}^{T} E\,(e_{it})^2$ and $V$ is $min_{\Lambda, F^k} (NT)^{-1} \sum_{i=1}^{N} \sum_{t=1}^{T} \left(X_{it} - \lambda_i^k F_i^k\right)^2$ (Bai & Ng, 2002, pp. 192, 197, 198, 201). Under such criteria (IC2), were selected two common factors ($r = 2$), and the VAR lag length is selected on the estimated factors with 'VARselect' from 'vars' library (AIC, $lag.max = 4$), yielding $p = 1$.

To achieve a delimitation of the productive core that is ontologically consistent with value theory yet statistically robust against data complexity, a multi-stage econometric architecture centered on Dynamic Factor Models (DFM) was designed and implemented. The analysis began with the rigorous pre-processing of the 47 sectoral time series of Gross Operating Surplus, which were transformed via logarithmic differences to ensure stationarity in both mean and variance, thereby eliminating stochastic trends that could lead to spurious regressions. Subsequently, these stationary series were standardized using *Z-scores*, a critical step to neutralize absolute scale differences between sectors and ensure that the extraction of latent factors responded to cyclical co-movement rather than the mere accounting magnitude of the largest branches. Upon this depurated base, the factor extraction algorithm was executed, determining the dimensionality of the latent space ($r$) via the Bai and Ng information criteria (specifically IC2), optimized for panels of moderate dimension, and establishing the dynamic lag structure ($p$) of the underlying vector autoregressive process via the Akaike Information Criterion (AIC).

**Figure 12**

*Determination of the Optimal Number of Factors*

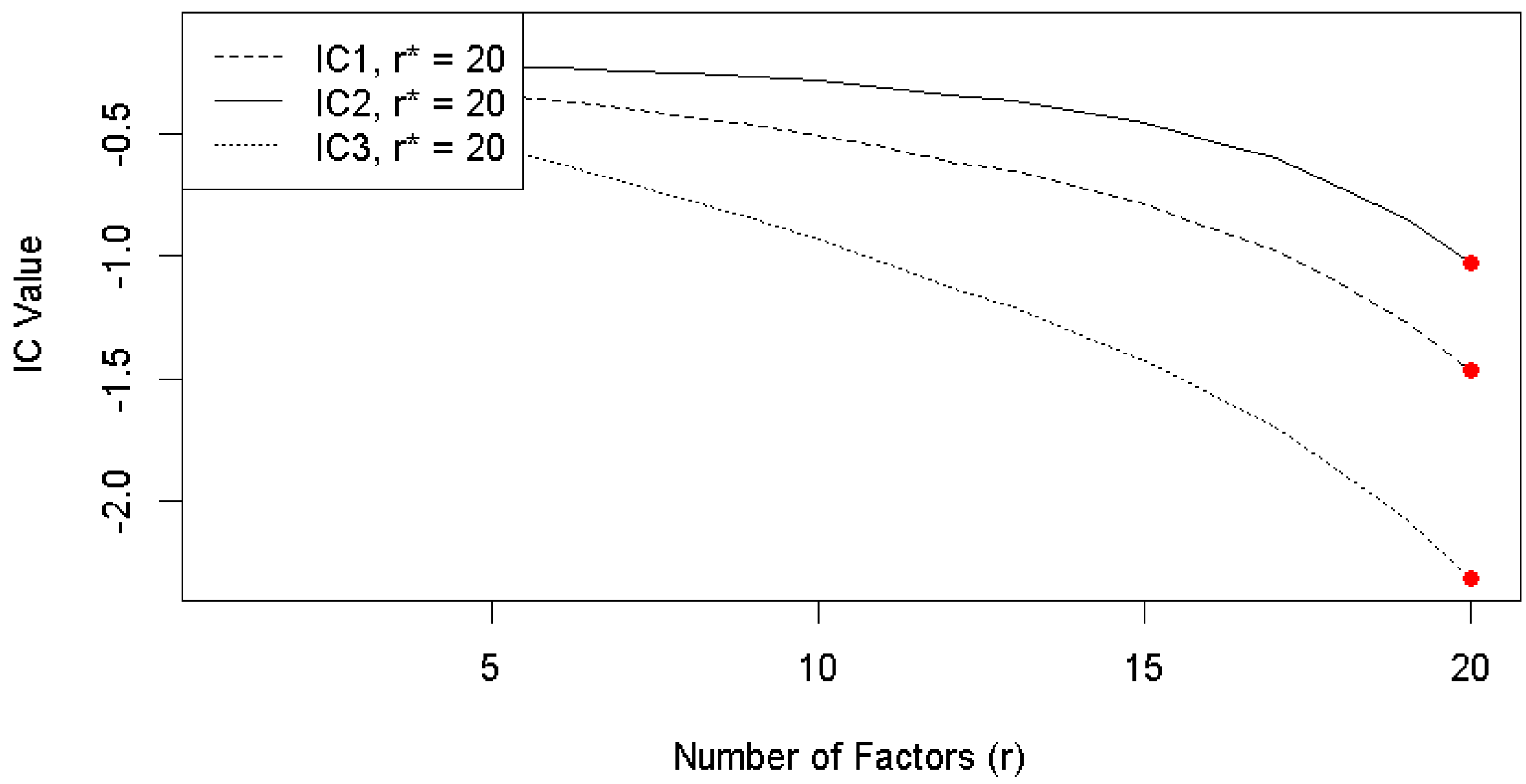

To disentangle mixed signals and endow the abstract components with economic interpretability, a Varimax orthogonal rotation was applied to the factor loading matrix, which allowed for a sharp segregation of the variance associated with structural persistence (trend) from that linked to transitory shocks (cycle).

Upon the identified factorial structure, a multi-horizon predictive validation scheme ($h = 1$ to 4 years) was deployed, designed to test the causal relevance of each sector under different temporalities of accumulation. For each prediction horizon, a high-dimensional design matrix was constructed, integrating both the common latent factors and the distributed lags of each individual sector. Given that the simultaneous inclusion of multiple lags per sector generated an over-parameterization problem, a penalized Elastic Net regression with a mixing parameter $\alpha = 0.5$ was employed, balancing the variable selection of LASSO with the collinearity management of Ridge. However, to avoid the instability inherent in variable selection within finite samples, this procedure was reinforced by a Stability Selection algorithm: 200 random subsampling iterations were executed on the data, retaining as valid predictors only those sectors that survived the selection process in more than 70% of the sub-samples. This severe filter ensured that the resulting "core" was not an artifact of outliers, but a persistent structural property of the data.

The definitive hierarchization of sectoral relevance was constructed by triangulating two orthogonal metrics derived from the validated model: idiosyncratic predictive capacity and the contribution to systemic transmission. For the former, the Partial $R^2$ was calculated, isolating the portion of the Aggregate Average Rate of Profit's variance that is explained exclusively by the internal dynamics of each sector, once the influence of global factors is discounted. For the latter, the interaction $|\lambda \cdot \beta|$ was computed, which weights the sector's factor loading (its coupling to the common cycle) by the sensitivity of the profit rate to said factor, thus identifying those nodes acting as "transmission belts" of the general cycle.

Statistical inference on these estimators was performed using a Synchronized Block Bootstrap ($B = 300$ replications), an advanced technique that resamples contiguous time blocks simultaneously for all variables; this allowed for the construction of empirical coefficient distributions while preserving the structure of serial autocorrelation and cross-sectional dependence intact, thereby overcoming the limitations of traditional asymptotic hypothesis tests that assume independence or normality in errors.

Finally, to shield the results against the rotational indeterminacy inherent in factor models and confirm the significance of the factor loadings, an adaptive thresholding algorithm termed "Full-Robust Thresholding" was developed. This method generated counterfactual null distributions via a block bootstrap on the original data, breaking any genuine structural relationship while maintaining the statistical properties of the series. To make the factor loadings of these null samples comparable to those of the original model, a Procrustes Alignment assisted by the Hungarian algorithm was implemented, correcting for any sign inversion or factor permutation that occurred during resampling. Cutoff thresholds were established based on the 95th percentile of these aligned null distributions, both at the specific factor level and within a global pool. Only those sectors whose empirical factor loadings exceeded this maximum noise threshold were considered valid components of the productive core, ensuring that the final selection of determinant sectors—subsequently used to filter the profit rate trend—rested on unshakable statistical evidence.

The specification of the Dynamic Factor Model (DFM) was rigorously grounded in the use of sectoral contributions to Gross Operating Surplus (GOS) as the input vector, deliberately discarding the use of net profits to preserve the ontological integrity of the Marxist analysis. This methodological decision responds to the critical need to isolate the dynamics of surplus value generation ($S$) in the productive sphere, thus avoiding contamination introduced by distributive noise—taxes, rents, interest, and transfers of value—inherent to the sphere of circulation and financial appropriation. By extracting the common latent factor from the GOS series, the model successfully identifies the underlying metabolic cycle of real accumula-tion, selecting exclusively those sectors whose factor loadings reveal a high structural synchronization with the systemic pulse of valorization. Consequently, the subsequent construction of the Average Rate of Profit (ARoP) upon this "generating core" ensures that the secular declining trend isolated later by the filtering algorithms is not an artifact of income distribution regimes, but the pure manifestation of the laws of capital production operating upon the organic composition of the productive core.

Furthermore, the specification of the Dynamic Factor Model for delimiting the productive core was based strictly on the sectoral Gross Rate of Profit, a methodological choice oriented towards capturing the metabolic intensity of accumulation devoid of scale or accounting distortions. By prioritizing an intensity variable (rate) over a magnitude variable (mass), the inertial bias towards merely voluminous sectors is neutralized, forcing the algorithm to identify those nodes whose profitability dictates the systemic pulse of valorization; simultaneously, the use of the gross rate isolates the generation of surplus value at its immediate productive source, preventing the extraction of the common signal from being contaminated by exogenous depreciation rules that affect the net rate, thereby guaranteeing that sectoral selection responds exclusively to the dynamics of capital efficiency rather than the administrative management of its wear and tear.

**Figure 13**

*Empirical Distribution Fitting of Sectoral Contributions to Principal Components via Maximum Goodness-of-Fit (MGE)*

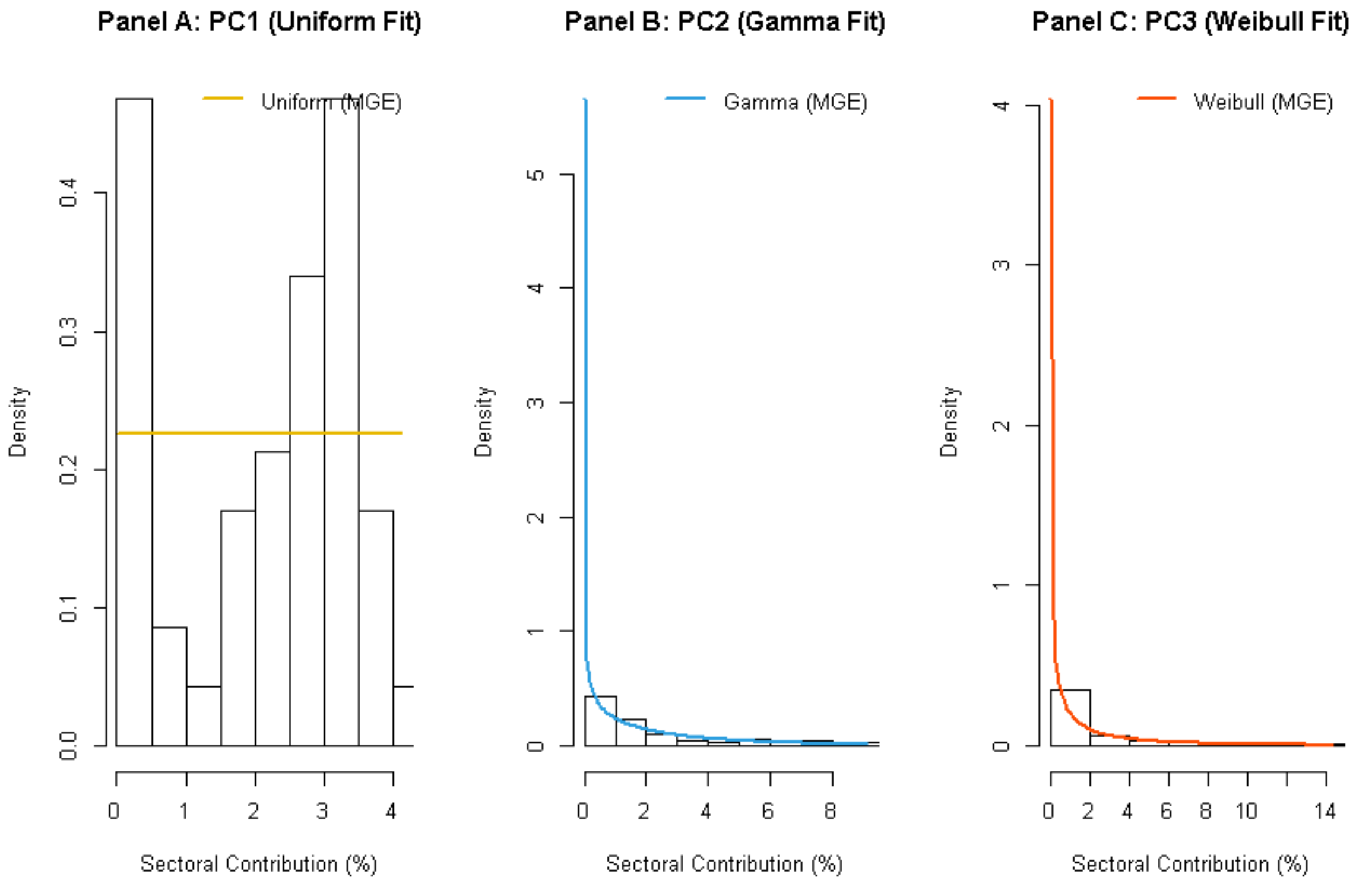

To rigorously validate the sector selection threshold, the empirical distribution of sectoral contributions to the first three principal components was modeled using the Maximum Goodness-of-Fit (MGE) method based on the Cramér-von Mises statistic. As illustrated in **Figure 13**, the first component (PC1) exhibits a Uniform distribution **(Panel A)**, suggesting a generalized variance structure potentially driven by aggregate size effects rather than structural differentiation. In contrast, the subsequent dominant components (PC2 and PC3) strictly adhere to Gamma **(Panel B)** and Weibull **(Panel C)** distributions, respectively. These heavy-tailed distributions statistically confirm the existence of a hierarchical 'core-periphery' topology within the productive matrix, where a small subset of sectors disproportionately drives the systemic variance, thereby providing a robust statistical mandate for the 'Top 10%' selection criterion employed.

The Dynamic Factor Model (DFM) estimation, specified with $r = 2$ factors and $p = 1$ lag based on the Bai–Ng IC2 and AIC criteria, achieved convergence in 26 iterations of the EM/Kalman algorithm, indicating robust identification and numerical stability. As illustrated in Figure 14, the estimated factors (colored lines) successfully track the common co-movement underlying the noisy aggregate of sectoral variances.

**Figure 14**

*Signal Extraction: Estimated Common Factors vs Sectoral Idiosyncratic Dynamics*

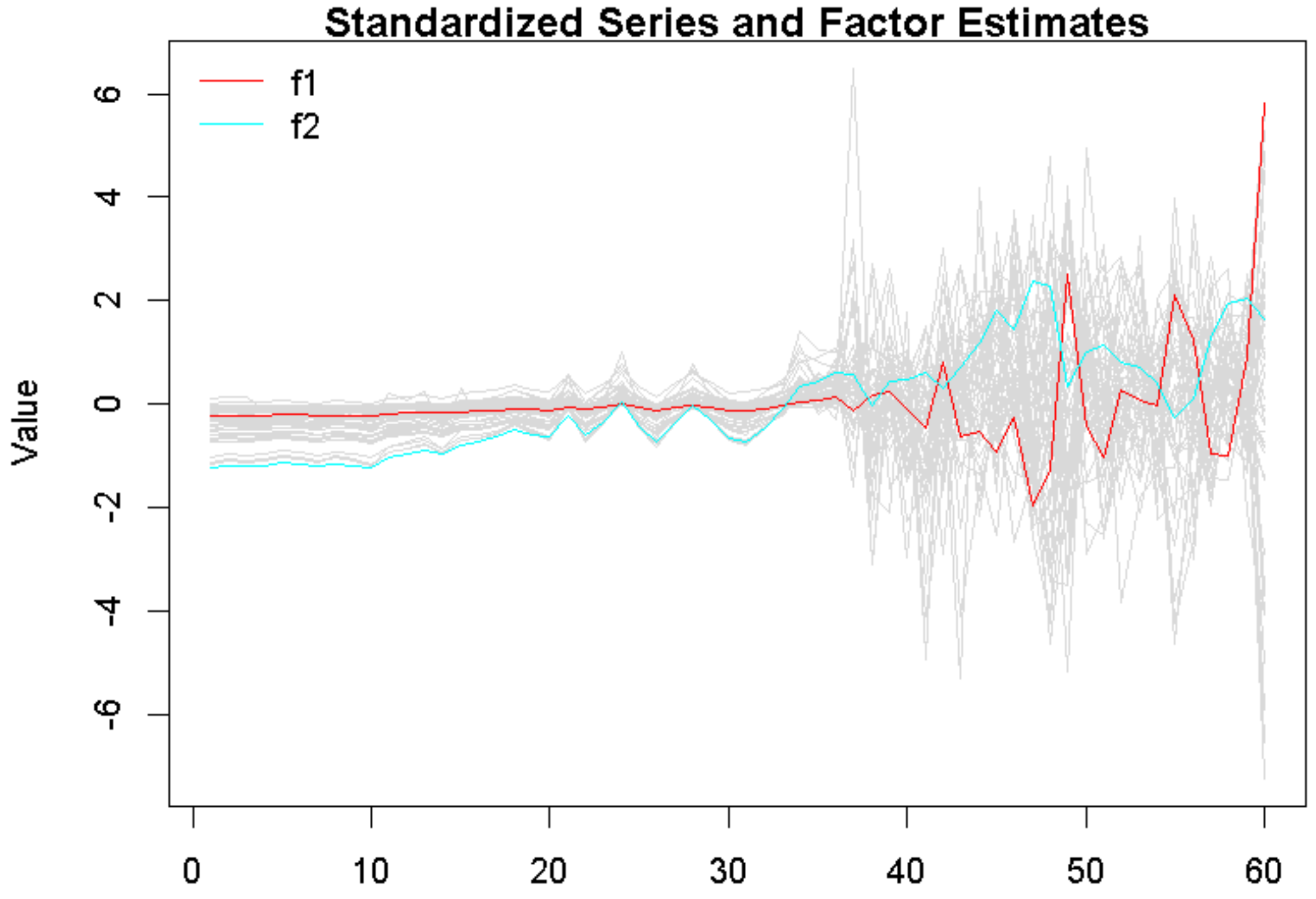

The estimated transition matrix, $A = [[0.3329,0.3253],[0.1680,0.9129]]$, reveals a distinct structural dichotomy in the latent dynamics of the US economy: the first factor exhibits low persistence (0.33), capturing short-term cyclical shocks, while the second factor displays high persistence (0.91), effectively acting as the carrier of the secular trend and long-term accumulation regimes.

**Figure 15**

*Latent Factor Dynamics and Persistence Regimes of the United States Economy (1960-2020)*

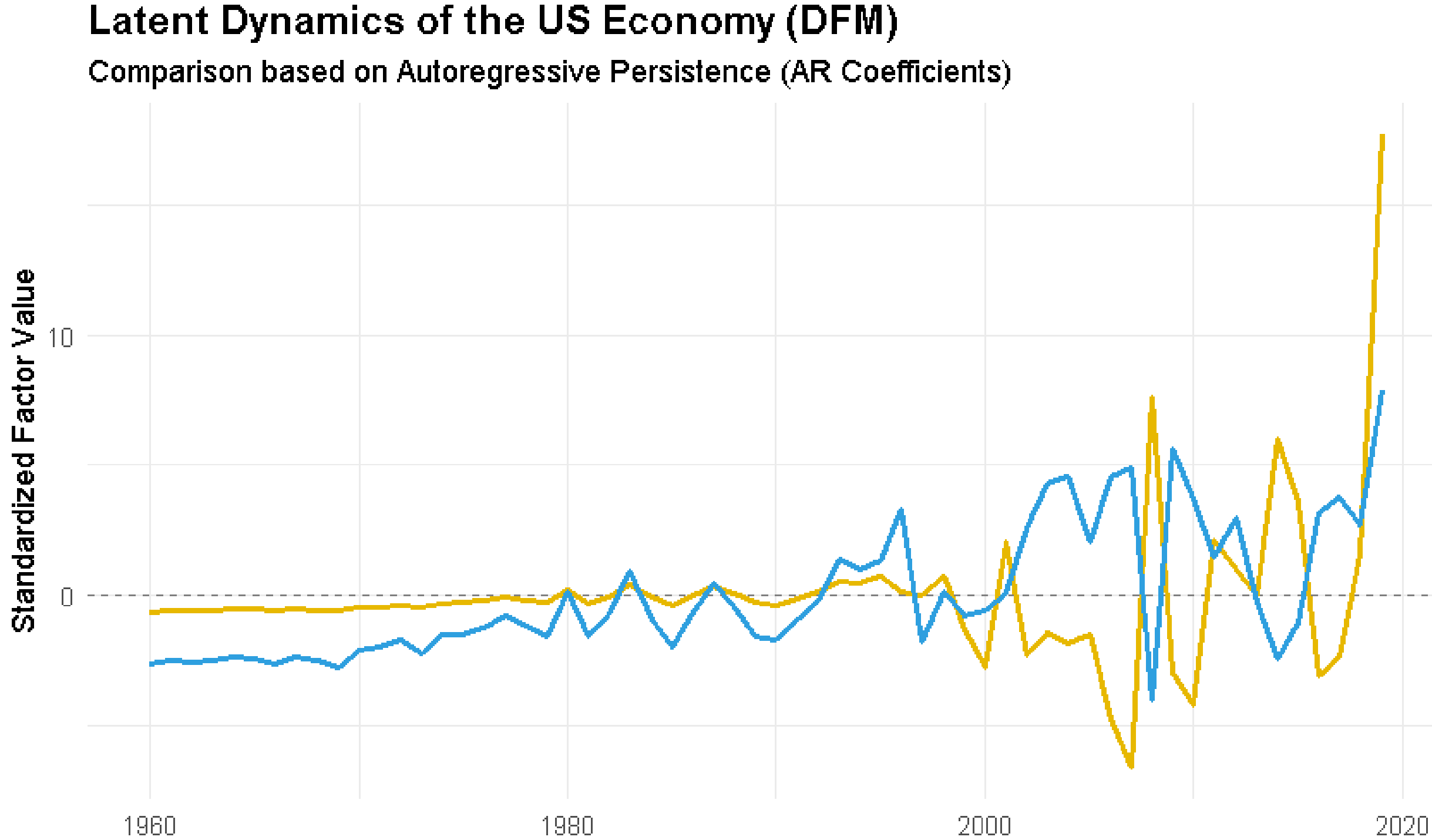

In terms of global goodness-of-fit, the two extracted common factors account for approximately 34.01% of the total variation in the standardized sectoral panel, as illustrated by the sharp decay in the scree plot (see Figure below). This level of systemic synchronization is corroborated by an average communality (mean $R^2$) of 0.3190. While these figures confirm the existence of a common metabolic pulse, they also highlight the significant heterogeneity of the US economy, where a substantial portion of sectoral profitability dynamics remains driven by idiosyncratic, branch-specific shocks rather than the global factor structure.

**Figure 16**

*Scree Plot of Sectoral Variance Explained by Principal Components*

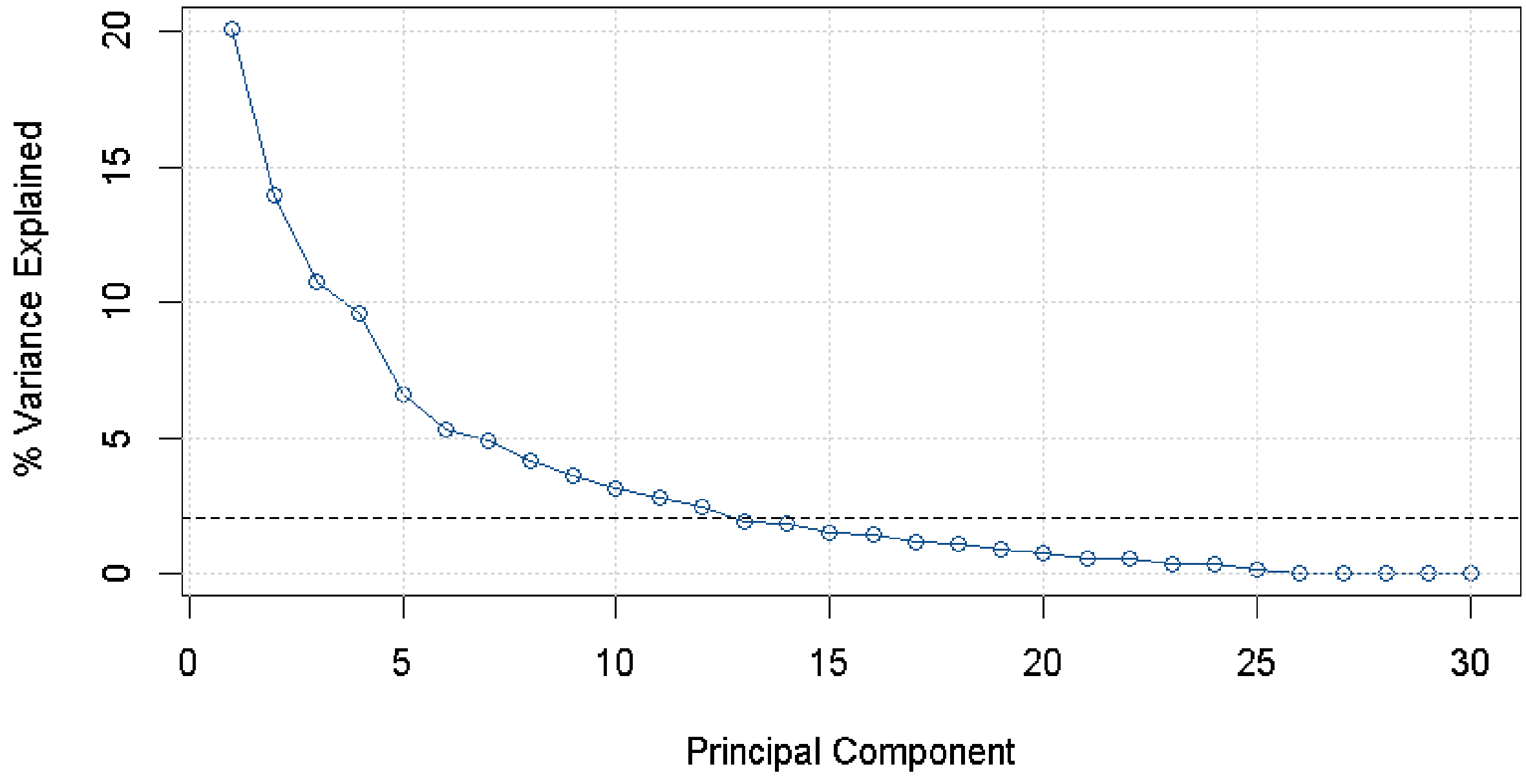

Regarding the specific factor content and predictive value, the rigorous application of our “four-pillar” anti-spurious screen (conditional Wald test with HAC bootstrap, pre-whitened directed CCF, placebo test, and CUSUM stability with FDR control) yielded an empty set of individually predictive sectors. Far from being a failure of the model, this result empirically validates the hypothesis of ontological interdependence: no single sector possesses the statistical strength to unilaterally predict the Aggregate Rate of Profit without failing strict stability tests. The profit rate is thus confirmed as a systemic class phenomenon, where the signal emerges from the collective interaction of the core rather than from isolated “silver bullet” sectors.

Despite this systemic distribution, the intensity of synchronization is hierarchically structured. The empirical distribution of the sectoral coefficients of determination (R2)—which measure the coupling of each sector to the common cycle—provides the best fit to a Weibull distribution (BIC = −6.41; shape $k = 1.15$, scale $\lambda = 0.36$), significantly outperforming the Normal and Gamma alternatives. A shape parameter near unity (1.15) implies a distribution resembling the exponential family, indicating a core-periphery structure where most sectors operate with idiosyncratic noise (low $R^2$), while the systemic signal is concentrated in a heavy right tail of highly integrated nodes.

**Figure 17**

Godness-of-Fit Diagnostics for the Sectoral Syncrhonization Distribution (Weibull Fit of $R^2$)

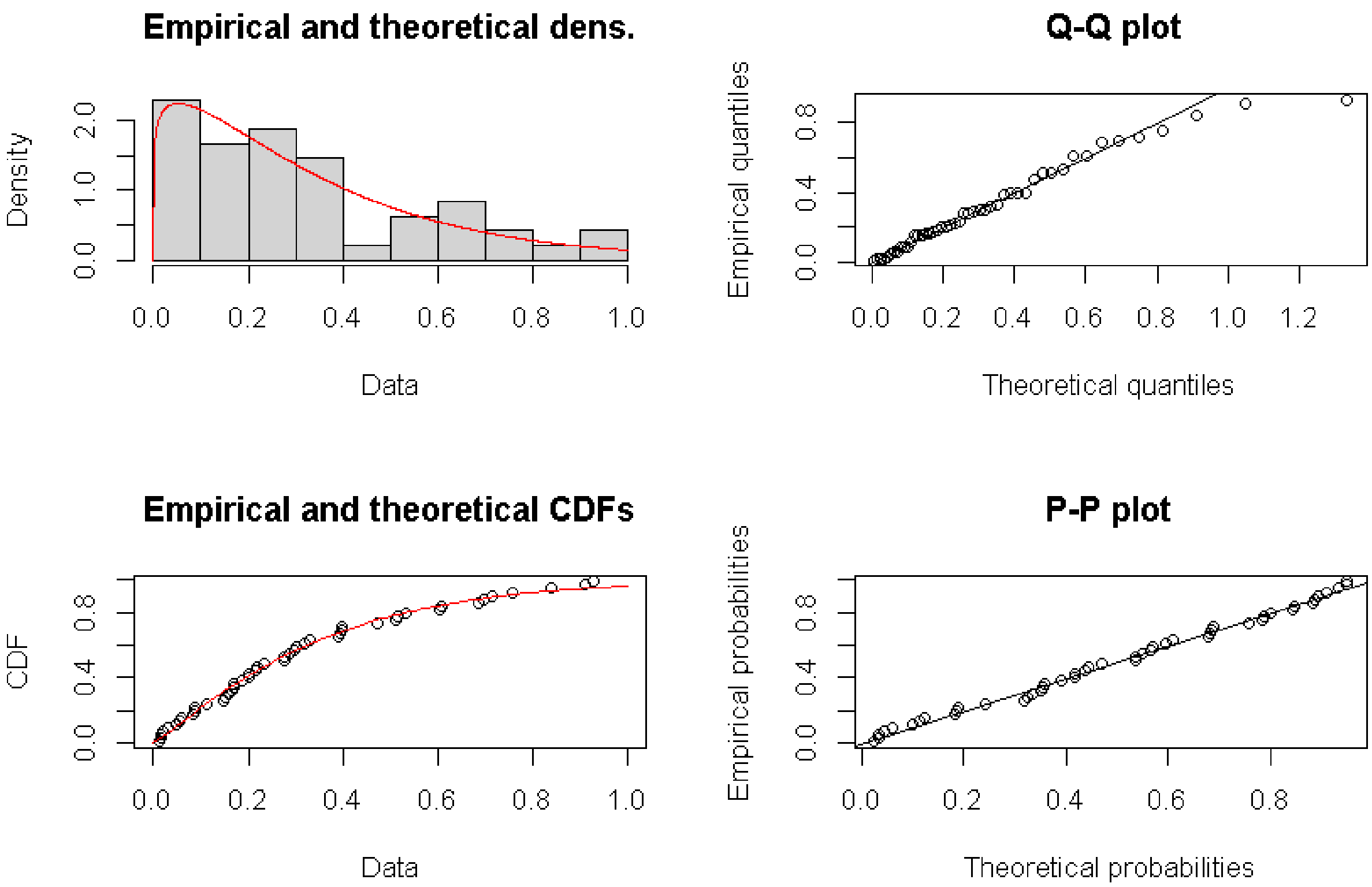

By triangulating the robust factor loadings (via Full-Robust Thresholding with synchronized bootstrap) and the multi-horizon predictive scores, we identified this “empirical core” which drives the aggregate profitability trend. Contrary to the expectations of a purely industrial value theory, the sectors with the highest structural weight are Real Estate (V33), State and Local General Government (V46), and Federal General Government (V44), followed by cyclical drivers in Retail Trade (V28) and Food Services (V42), with Utilities (V6) and Chemical Products (V25) providing the industrial baseline.

**Figure 18**

*Hierarchical Structure of the Productive Core: Sectoral Factor Loadings*

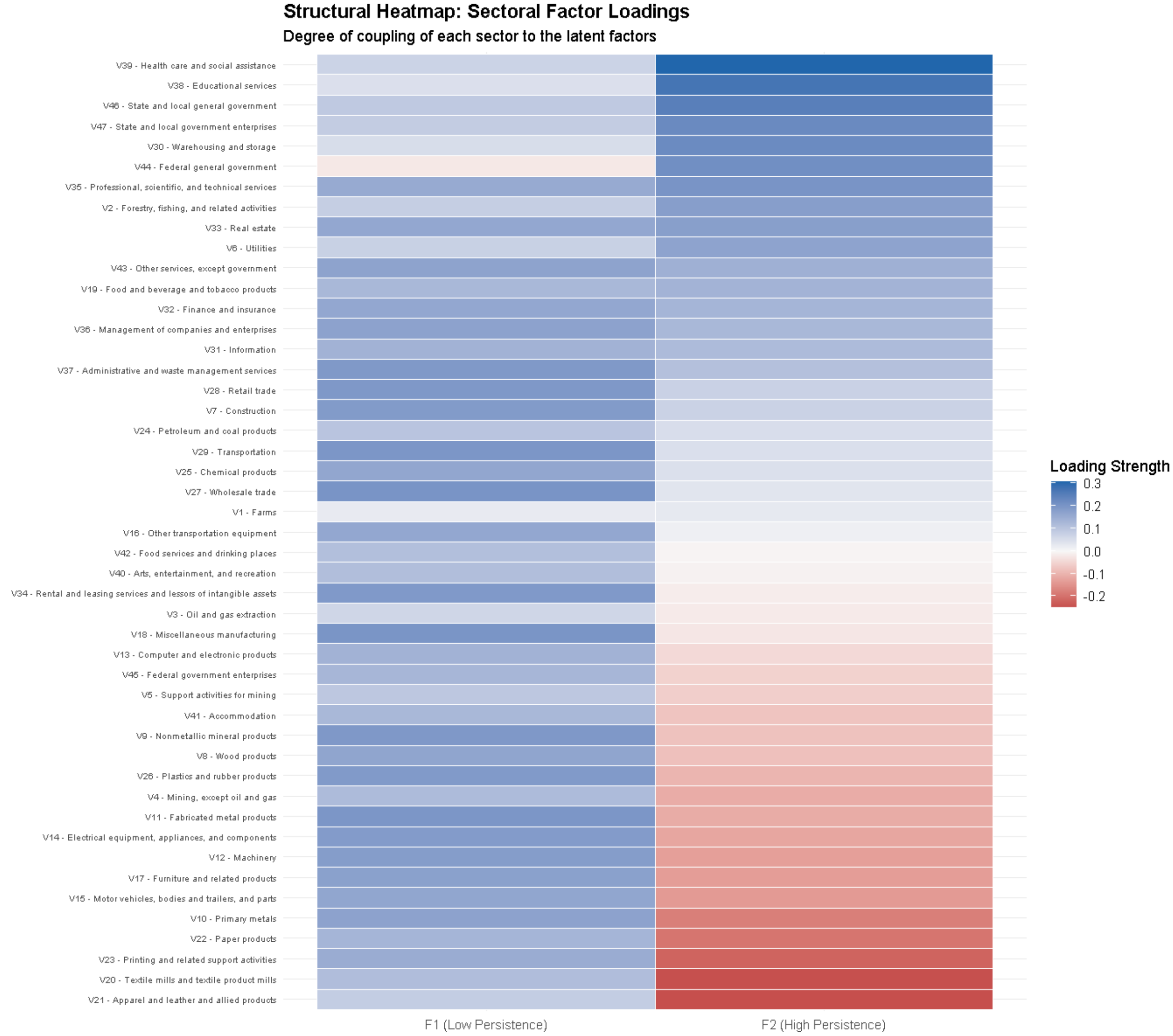

This composition suggests that the empirically observed "mass of profit" in National Accounts is heavily dominated by the inertia of imputed rents and public fixed capital consumption. While these sectors are theoretically classified as unproductive or revenue-consuming within the Marxist framework, their statistical dominance in the DFM confirms that they have effectively colonized the macro-dynamics of the US rate of profit, acting as the primary gravitational force of its secular trend during the 1960–2020 period.

### III.III.IV. Results of Trend Filtering for Marxist ARoP Using PCA, RHR, and DFM Criteria

**Figure 19**

*Sectoral Composition of the Average Net Profit Rate According to Principal Component Analysis: Secular Trends Obtained by Different Filtering Methods*

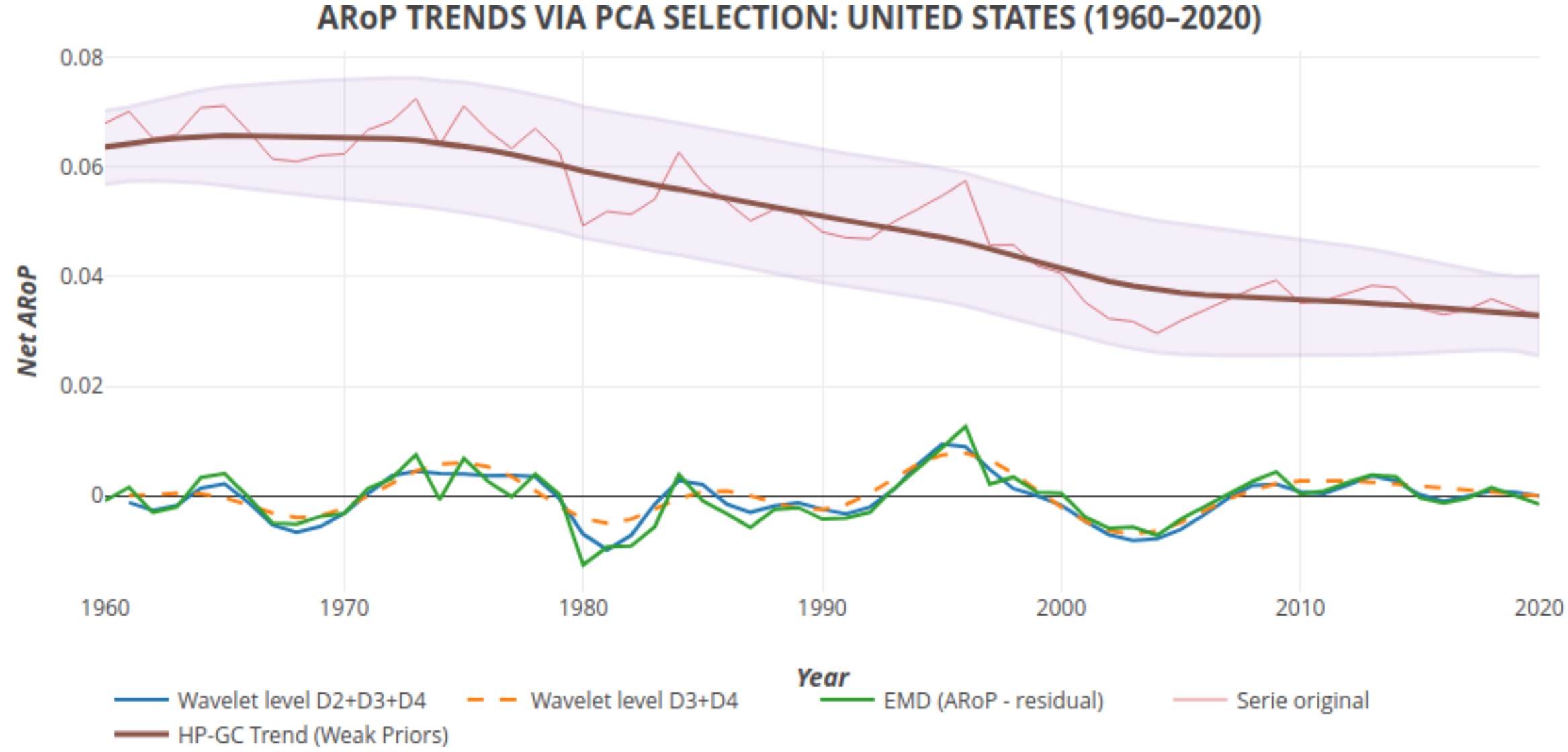

**Figure 20**

*Sectoral Composition of the Average Net Profit Rate According to Regularized Horseshoe Regression: Secular Trends Obtained by Different Filtering Methods*

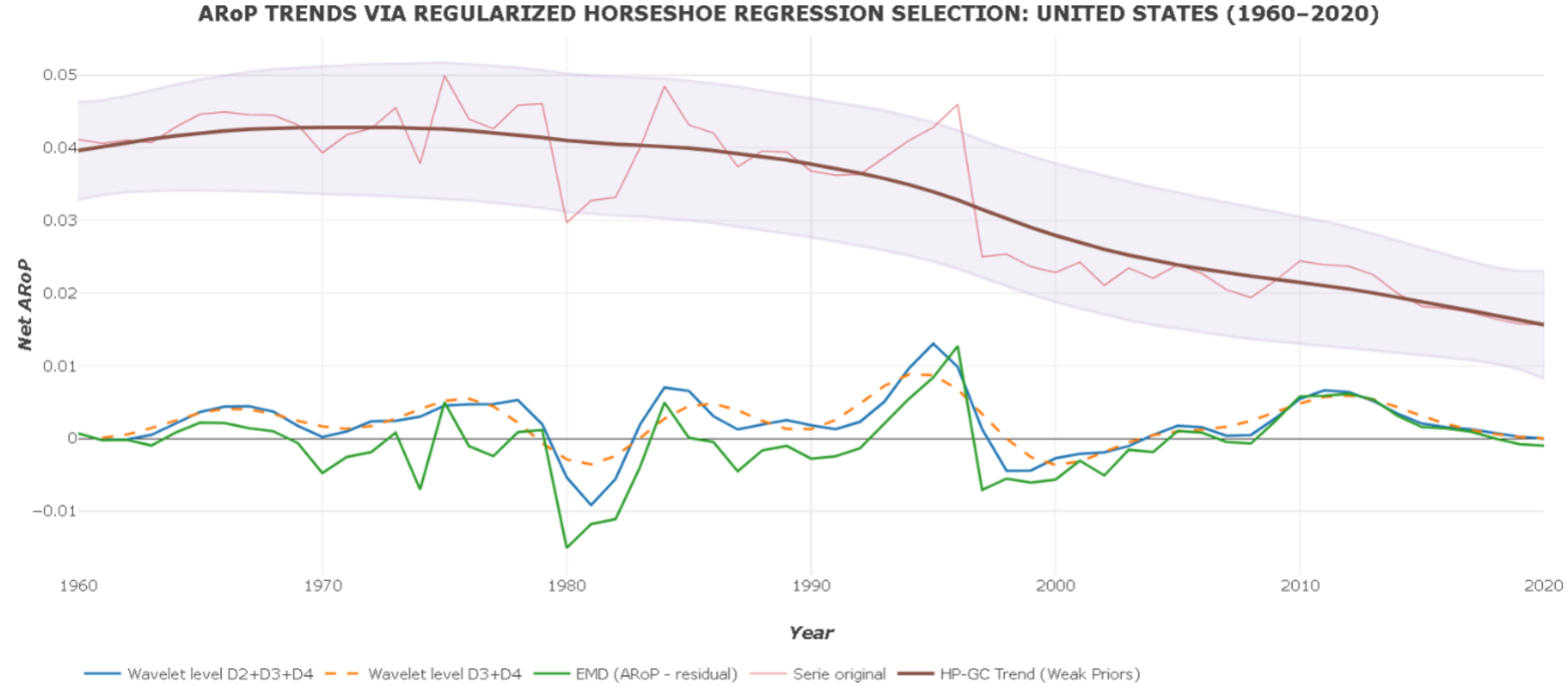

**Figure 21**

*Sectoral Composition of the Average Net Profit Rate According to Dynamic Factor Modeling: Secular Trends Obtained by Different Filtering Methods*

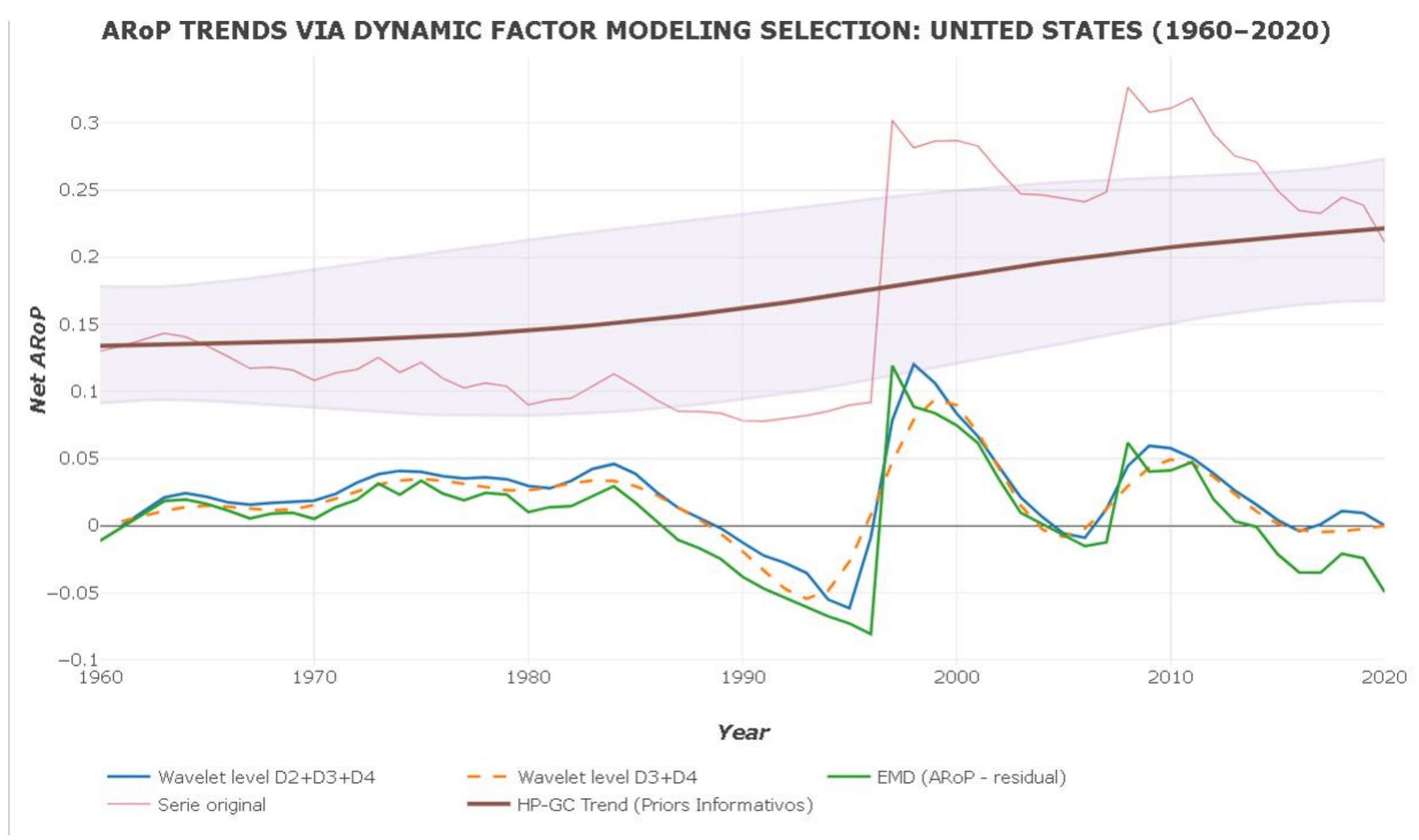

With the very exception of the HP filter under DFM criteria, all the obtained results are qualitatively consistent with the central conclusion obtained with the constructed criteria: the Marxist net ARoP falls in the long term regardless of the filter applied, considering any key sectors, if there is robust theoretical and/or econometric work behind that consideration.

On HP filter some issues must be pointed out. This filter is linear in the sense that it aims to separate the time series into a trend component and a cyclical component through a linear combination of the original data. In this regard, it is a parametric technique because it assumes a specific structure for modeling the trend and cyclical component. Additionally, this filter requires the specification of several parameters, such as the desired smoothness of the trend and the cyclical component, which in our research were defined as the variances of the trend and the cyclical component itself, respectively. Furthermore, the HP filter uses a model structure that assumes a second-order Markov process for the trend, which is also a parametric specification.

The previous issues imply that the filter may be affected by the sample size (even when the Gibbs sampler has been used) and/or because the specified distributions have not been appropriate. The EMD filter was the one in which the downward trend of the Marxist net ARoP was most accentuated, which has the advantage of being a non-parametric technique that can adapt well to different patterns in the data, which makes it useful for identifying underlying components of complex time series, dispensing with prior assumptions about the functional form of the time series.

## IV. CONCLUSIONS

The present research successfully established and validated a standard criterion grounded in Marxist theory for the inclusion and exclusion of economic activities in the calculation of the Average Rate of Profit (ARoP). The theoretical criteria were systematically constructed in Section II.II and empirically applied to the United States economy (1960-2020), starting from the fundamental distinction between productive and unproductive labor, the sectoral location in the circuit of capital, and the organic relationship with the production of surplus value.

The validation of these criteria was performed through two fundamental methodological dimensions: internal consistency (theoretical) and external consistency (objective/econometric).

### IV.I. INTERNAL CONSISTENCY AND THEORETICAL VALIDATION

The operational application of the proposed criteria involved a minute contrast with the North American Industry Classification System (2022). The theoretical pertinence of including mixed sectors—those amalgamating productive and unproductive functions—under specific conditions was concluded.

In particular, sectors such as **Educational Services (V38)** and **Administrative and Waste Management Services (V37)** were included. Although they contain components that are not value-generating (such as non-market public education or security and cleaning services), their inclusion was determined because education is fundamental for the reproduction of the skilled labor force in an industrialized economy, and administrative services include the preparation of documents and management for productive sectors, assuming that the majority of these services are integrated into the productive capital circuit. Likewise, the **Warehousing and Storage (V30)** sector was included, given that the preservation of the use-properties of merchandise is a continuation of the production process within the sphere of circulation, indispensable for the very existence of value.

Conversely, other sectors were rigorously excluded which, although they might contain tangential activities of transportation or conservation, operate systematically under logics alien to the production of surplus value. **Trade (V27, V28)** and **Finance (V32)** were rigorously excluded for belonging to the sphere of pure circulation, where metamorphoses of value (buying-selling) are realized without the creation of new value. Similarly, the **Government (V44, V46)** was excluded for not being governed by profit maximization, as were the **Real Estate (V33)** and **Health Care and Social Assistance (V39)** sectors, given that their predominant logic is the appropriation of rents or the provision of non-market services that do not objectify abstract labor generating capital.

To verify the internal consistency with the Law of the Tendency of the Rate of Profit to Fall, the time series of the ARoP was subjected to state-of-the-art filters. The results demonstrated a secular declining trend consistent with Marx's predictions. Additionally, an exhaustive battery of 17 unit root tests—including variations of ADF, Elliott-Rothenberg-Stock (ERS), KPSS, and Phillips-Perron—confirmed, across 35 distinct specifications, that the series is non-stationary in levels but stationary around a trend. This consensus among tests of varying power and specification (with and without drift, with and without deterministic trend) constitutes strong statistical evidence that profitability does not follow a random walk where shocks are permanent and direction is stochastic. On the contrary, the results indicate that the trajectory of the rate of profit is governed by a **deterministic underlying trend**; that is, there exists a structural force (the law of value) that imposes a long-term gravitational direction, which persists beyond cyclical fluctuations or transitory market shocks.

## IV.II. EXTERNAL CONSISTENCY: ADVANCED ECONOMETRIC EVIDENCE

The objective robustness was subjected to rigorous econometric scrutiny using three methodologies agnostic to Marxist theory, the results of which validate the hypothesis of structural interdependence.

### IV.II.I. Principal Component Analysis (PCA) and the Center-Periphery Topology

The analysis of the variance structure revealed a clear statistical hierarchy. While the first principal component (PC1) explains the largest proportion of variance (47.3%), its empirical distribution is Uniform, suggesting that it captures the general "size effect" or the aggregate inertial growth of the economy, without discriminating structure. In contrast, the subsequent components (PC2, PC3), which explain structural differentiation, strictly fit heavy-tailed distributions (Gamma and Weibull). This confirms that the qualitative systemic variance is not driven by the uniform aggregate, but by a small subset of sectors (Top 15%) located in the tails of these distributions. The exclusion of irrelevant sectors under this criterion resulted in a productive matrix highly correlated with the theoretical Marxist criteria.

When we speak of 'qualitative systemic variance', we seek to differentiate the structure of variance from its mere magnitude. While the First Principal Component (PC1) captures the 'size effect' or inertial growth (uniform quantitative variance), the subsequent components (PC2, PC3) capture the functional and hierarchical organization of the economy. It is termed 'qualitative' because the fit to heavy-tailed distributions (Gamma/Weibull) in these components reveals *who* generates the dynamics (the identity of the determinant sectors) and not just *how much* the economy varies, allowing for the isolation of the structural signal from the background noise.

The classification of PC1 as 'background noise' or 'aggregate inertia' is not a dismissal of its quantitative magnitude, but a characterization of its null informational content for the purposes of structural discrimination. From the perspective of Information Theory (Shannon Entropy), a Uniform distribution represents the state of maximum entropy (maximum uncertainty) within a bounded domain; in the context of PCA, this means that the loading vector of PC1 contains no information allowing for the distinction of the relative importance of one sector over another, as all sectors 'load' on the general movement of the economy in an undifferentiated manner.

In the multivariate analysis of complex systems, this first principal component isolates precisely the size effect or the non-discriminating common trend—aggregate inertial growth and generalized inflation—that affects all sectors equally (maximum entropy). It functions, therefore, as that 'background noise' or 'tide' that lifts all nominal magnitudes without revealing the internal causal structure. Economically, this captures the inertia of the size of the economy (where 'a rising tide lifts all boats') and, fundamentally, the **Brownian motion of the system**.

Conceptualizing this phenomenon as 'Brownian motion' is rigorous from a dialectical standpoint, as it captures the anarchy of production in its phenomenal manifestation. From the perspective of classical Marxism and dialectical materialism, this Brownian motion of the system must be conceived as the dialectical unity between chance and necessity, where the apparent randomness and stochastic chaos of prices, generalized inflation, and the inertia of aggregate growth (captured statistically by PC1) are not mere external accidents, but the necessary form of appearance in which the law of value is realized.

The multiple idiosyncratic shocks and the nominal volatility of individual capitals act here as the 'particles' whose erratic and blind movement, when aggregated socially, ends up violently imposing the hidden regulation of socially necessary labor time as a natural regulating law that operates 'behind the backs of the producers'. In this way, the sum of stochastic movements crystallizes into an inertial macroscopic regularity which, although dominating quantitative variance, lacks the qualitative information necessary to identify the true structural motors of accumulation and crisis. Thus, it is revealed that this superficial 'noise' is, in reality, the turbulent and immanent mechanism through which the organic totality of capital reproduces its conditions of existence, socially validates private labor, and unfolds its structural contradictions.

For the objective of this research—which is to identify a causal or determinant core—a signal that affects all variables equally is, by definition, noise, as it lacks discriminant power; it does not reveal the internal structure of the mechanism of accumulation, but only its scale. In contrast, the Gamma and Weibull distributions of the subsequent components (PC2, PC3) represent states of low entropy, where variance is concentrated ('collapsed') in specific nodes. This statistical asymmetry is what constitutes the structural signal: it is the fingerprint of functional differentiation, allowing the mathematical isolation of the sectors acting as motors of the cycle from those that are mere passengers of the aggregate trend.

### IV.II.II. Regularized Horseshoe Regression and Value Causality

The Regularized Horseshoe Regression (RHR) validated the conception of the economy as an organic totality, where multicollinearity is an ontological characteristic. Despite the collapse of coefficients due to this interdependence, the predictive projection ranking (*projpred*) identified a productive core led by Retail Trade, Textiles, and Fabricated Metal Products. This alignment with the notion of determinant sectors in the creation of value is justified because the model was specified using total **Variable Capital** as the predictor of Gross Surplus. Under the Labor Theory of Value, variable capital is the sole source of surplus value; therefore, the sectors that the algorithm selects as the best predictors (even under strong regularization) are those where the functional relation (deployment of labor power generation of surplus) is most robust and systemically determinant, filtering out those where surplus comes from transfers or rents.

### IV.II.III. Dynamic Factor Models (DFM) and the Metabolic Pulse

The estimation using DFM confirmed that the economy is not a sum of disjointed parts, but a synchronized system. The model revealed that only two latent factors explain **34.01%** of the total variance of 47 sectors, a significant figure given that the work is done with differenced (stationary) data. Even more compelling is that the distribution of the sectoral coefficients of determination ($R^2$) fits a **Weibull** distribution, which mathematically demonstrates that there exists a common "pulse" or metabolic cycle orchestrating profitability: the majority of sectors have a weak connection (idiosyncratic noise), but there exists a "right tail" of highly integrated sectors that transmit the general dynamic of accumulation.

### IV.II.IV. The Dialectic Between Essence and Appearance: Horseshoe vs. PCA and DFM

A fundamental finding lies in the phenomenological divergence between the identified cores. While the **Horseshoe** (causal logic $v \rightarrow s$) prioritized tradable sectors, the **DFM** (logic of common variance) and the **PCA** assigned dominant weights to unproductive sectors, albeit with notable differences in their governmental composition:

- The **DFM** identified **Real Estate (V33)** and a strong state presence with the **Federal General Government (V44)** and **State and Local General Government (V46)** as central.
- The **PCA**, for its part, highlighted the **Federal General Government (V44)** and its enterprises (V45), but explicitly excluded the **State and Local General Government (V46)** from its main core.

This divergence validates the duality of capital: the Horseshoe identifies the **essence** (value generation), while DFM and PCA capture the **appearance** (an economy phenomenologically dominated by real estate rents and public spending). However, the crisis of profitability is so deep that it manifests in both cores.

### IV.II.V. Trend Analysis and the Anomaly of the HP-GC Filter

Synthesizing the trajectories of the Average Rate of Profit, it is observed that the declining trend is not a monotonic straight line, but a dialectical process. The non-parametric filters (**EMD** and **Wavelets**) reveal a marked but oscillatory decline, evidencing the constant struggle between the general tendency toward decline and the counteracting mechanisms (devaluation shocks, bubbles) that attempt to momentarily reverse it.

However, a technical anomaly was detected in the series constructed via **PCA** and **DFM**: the **HP-GC** filter projects a flat or slightly rising trend. This discrepancy is an artifact of the filter's rigidity in the face of the nature of the data. PCA and DFM include rentier sectors (such as Real Estate) whose "profits" (imputed rents) grow in an inertial and constant manner, without the cyclical volatility of real production. This accentuation of the trend in the case of the HP-GC filter is not a random artifact but a direct consequence of its state-space specification, which mathematically penalizes the variance of the trend component's second difference (*i.e.*, its acceleration). By imposing a 'global memory' structure on the series, the algorithm forces a smoothing process that connects the entire history of the data, thereby restricting the model's capacity to adapt to abrupt regime changes or local volatility. Consequently, this configuration tends to linearize the signal extraction process, masking the breaking points that non-parametric methods, sensitive to local frequency, do detect. Therefore, this "recovery" in HP-GC indicates not economic health, but rentier hypertrophy concealing productive atrophy.

### IV.III. FINAL SYNTHESIS

The research concludes by fully satisfying the criterion of the **Coherence Theory of Truth** posited in the introduction. The new proposition—the Marxist criteria for sectoral exclusion—has demonstrated not to be in contradiction with the system of existing propositions (the labor theory of value), but to bear a deep theoretical harmony with it. In turn, said harmony is sustained in terms of the objective evidence of the secular behavior of the average rate of profit, processed through advanced econometric algorithms.

Regardless of whether the sectoral core is defined by theory (productive labor), variance dominance (PCA), predictive capacity under uncertainty (Horseshoe), or latent synchronization (DFM), the Net Average Rate of Profit exhibits a historical downward gravity when methodological noise is adequately depurated. This confirms that the Law of the Tendency of the Rate of Profit to Fall is not an artifact of data selection, but the fundamental empirical regularity governing the metabolism of the US economy, expressing the essential historical gravity of said economy.

This scientifically validates the conceptual relevance and practical utility of the Marxist distinction between productive and unproductive labor for contemporary macroeconomic analysis. The validation resides in a compelling epistemological fact reflecting the ontology of the system: when the Average Rate of Profit is calculated without discriminating sectors (naive calculation), the declining tendency may remain hidden under the statistical noise of circulation and rent; however, upon applying the exclusion criteria grounded in value theory, the tendential law emerges with empirical sharpness. This demonstrates that the categorization of 'productive labor' is not an arbitrary abstraction, but a real ontological category that isolates the motor of accumulation. If the distinction were irrelevant, the depuration of unproductive sectors would have yielded random noise instead of a structural regularity consistent with theoretical prediction, thus confirming that the explanatory power of the model increases, rather than decreases, by adhering to the strict definitions established by Marx.